%% file: preprint.tex
\documentclass{emulateapj}

\newcommand{\nh}{N_{\rm H,22}}

\slugcomment{To be published in ApJ 2009 October, vol.\ 702}

\shorttitle{CCO in Cassiopeia A }
\shortauthors{Pavlov and Luna}

\begin{document}

\title{A dedicated Chandra ACIS observation of the 
central compact object in the Cassiopeia A supernova remnant}

\author{G.\ G.\ Pavlov}
\affil{Pennsylvania State University, 525 Davey Laboratory, University Park, PA 16802, USA; \tt{pavlov@astro.psu.edu}}
\and
\author{G.\ J.\ M.\ Luna}
\affil{Smithsonian Astrophysical Observatory, 60 Garden Street, Cambridge, 
MA 02138, USA; \tt{gluna@cfa.harvard.edu}}

\begin{abstract}
We present results of a recent {\sl Chandra} X-ray Observatory observation 
of the central compact object (CCO) in the supernova remnant Cassiopeia A. 
This observation was carried out in an instrumental configuration that 
combines a high spatial resolution
with a minimum spectral distortion, and it allowed us to search for pulsations 
with periods longer than $\approx 0.68$ s.
We found no evidence of extended emission associated with the CCO, nor
statistically significant pulsations (the $3\sigma$ upper limit on pulsed fraction
is about 16\%).
The fits of the CCO spectrum with the power-law model 
yield a large photon
index, $\Gamma\approx 5$, and a hydrogen column density larger than
that obtained from the SNR spectra. 
The fits with the blackbody model 
are statistically unacceptable. 
Better fits are provided by 
hydrogen or helium neutron star atmosphere models,
with the best-fit effective temperature $kT_{\rm eff}^\infty \approx 0.2$ keV,
but they require a small star's radius, 
$R=4$--5.5 km,
and a low mass, $M\lesssim 0.8 M_\odot$. A neutron star cannot have 
so small radius and mass,
but the observed emission might
 emerge from an atmosphere of a strange quark star.
More likely, the CCO 
could be a neutron star with a nonuniform surface temperature and a low
surface magnetic field (the so-called anti-magnetar), 
similar to three other CCOs for which upper limits
on period derivative have been established. The bolometric luminosity,
$L_{\rm bol}^\infty \sim 6\times 10^{33}$ erg s$^{-1}$, estimated
from the fits with the hydrogen atmosphere models, is consistent
with the standard neutron star cooling for the CCO age of 330 yr.
The origin of the surface temperature nonuniformity remains to be understood;
it might be caused by anisotropic heat conduction in the
neutron star crust with very strong toroidal magnetic fields.
\end{abstract}


\section{Introduction}
\label{intro}

Cassiopeia A (Cas A) 
is the famous remnant of a Type II supernova explosion that 
was apparently detected about 330 years ago 
\citep{aschworth}.
The compact X-ray source at
the center of Cas A was discovered in the first-light {\sl Chandra}
observation 
\citep{tananbaum99} and then
found in archival {\sl ROSAT} and {\sl Einstein} images \citep{aschenbach,pavlov1999}. 
 It is considered as
 a prototype of the so-called compact central objects (CCOs) in 
supernova remnants, 
which are possibly neutron stars (NSs) with properties
very different from those of radio pulsars \citep{pavlov2002}.

The analysis by \citet{pavlov2000} 
and \citet{chakra2001}
 of first {\sl Chandra}
observations of the Cas A CCO
with the ACIS and HRC detectors
 has shown that it is a pointlike source 
with a flux 
$F_X\approx (8$--$9)\times 10^{-13}$ ergs s$^{-1}$ cm$^{-2}$ 
in the 0.6--6 keV band.  
The source showed no periodicity, with an upper limit on the pulsed
fraction of 35\% for $P>20$ ms.
Its spectrum was found to be consistent with an absorbed blackbody (BB) model 
[$\nh\equiv N_{\rm H}/10^{22}\,{\rm cm}^{-2} \approx 0.8$,
$kT 
= 0.5$--0.6 keV, $R
= (0.3$--$0.5) D_{3.4}$ km, where $D_{3.4}
=D/3.4\,{\rm kpc}$ 
is the distance
scaled to the 
remnant's estimated distance; \citealp{reed}] 
or with a 
steep 
power-law (PL) model
($\nh\approx 
2$, photon index $\Gamma 
= 3$--4).

Based on the X-ray properties of the Cas A CCO, \citeauthor{pavlov2000} and 
\citeauthor{chakra2001} suggested that its thermal-like radiation 
might be interpreted as emitted from hot spots at the 
NS surface, similar to those seen in some young radio pulsars and accreting X-ray pulsars.
However, this source 
shows no indications of pulsar activity 
(such as radio or $\gamma$-radiation, or a 
pulsar wind nebula), the emitting region is unusually hot compared to radio pulsars,
and the source flux does not show variations 
expected for an accreting object. 
\citet{pavlov2000} and \citet{chakra2001} speculated that the 
unusual properties of this source might be caused by an extremely 
high magnetic field, $B\sim 10^{14}$--$10^{15}$ G,
similar to those of anomalous X-ray pulsars (AXPs) and soft $\gamma$-ray
repeaters (SGRs), commonly known as ``magnetars'',
but 
they concluded that more observations were required to 
understand the true nature of this CCO from the spectral
and timing analysis.

Since the discovery of the CCO, Cas A has been observed many times with 
{\sl Chandra}.  In addition to about 
90 short (1--10 ks) ACIS and HRC calibration observations, two 50 ks ACIS-S3 observations (2000 January and 2002 February), three 50 ks HRC observations 
(1999 December, 2000 October, and 2001 September), and a 70 ks ACIS/HETG observation (2001 May) 
have been carried out.  
Finally, Cas A was observed with ACIS-S3 in 2004 February -- May 
(9 pointings, 40--170 ks exposures) for a total exposure of about 1 Ms.
Unfortunately, the vast amount of data collected in those observations
has not added much to our understanding of the CCO.
The results of the 50 ks ACIS-S3 observation of 2000 January
 and 50 ks HRC-S observation of 2000 October have been reported
by \citet{murray2002}. 
They found a low-significance period
of 12 ms in the HRC-S data, which was not confirmed in the
HRC-S observation of 2001 September \citep{ransom2002},
and concluded that the spectrum was
consistent
with that reported by \citet{pavlov2000}
and \citet{chakra2001}. 
\citet{hwang2004}
presented beautiful images and spectra of the Cas A SNR
obtained in the 1 Ms observations, but,
regarding the CCO, they only mentioned
that the best single-component model is a BB
with 
$kT = 422\pm 6$ eV and $R=(0.83\pm 0.03)D_{3.4}$ km; that fit, however,
was formally unacceptable ($\chi_\nu^2=1.57$ for 
315 degrees of freedom 
[dof]).
The analysis of the ACIS observations 
of 1999--2004 did not show
statistically significant changes of the CCO flux 
 \citep{teter2004}
, virtually ruling out the possibility
that the CCO emission is due to accretion.
The spectral analysis of those data has shown that the fits with 
one-component
thermal models (BB or NS atmosphere)
 leave 
large residuals at high energies, 
$E\gtrsim 4$ keV,
suggesting that either the surface temperature is nonuniform or 
the emission at higher energies is of a nonthermal (e.g., magnetospheric)
origin.

However, the spectral 
fits of those data,
and the source properties inferred from those fits,
suffered from large systematic uncertainties 
because all the 
``bare-ACIS'' (no-grating) 
observations 
were taken in {\em full frame mode} (frame time 3.24 s), 
which resulted in a significant photon pileup\footnote{If two or more photons
interact with the same detection cell within the frame time, they are
registered as a single event. This effect is known as pileup (see \S6.15
of The {\sl Chandra} Proposers' Observatory Guide [POG], ver.\ 11, at 
\url{http://asc.harvard.edu/proposer/POG} for a detailed description).},
with $\sim 20$\% pileup fraction for plausible spectral models.  
Although such a pileup fraction
may look moderate,
it substantially distorts the spectrum.  In particular, it leads to 
an artificial excess of counts at higher energies 
(because two photons arriving within a frame are counted as one photon 
of higher energy), which can be easily confused with, e.g., a 
PL tail expected for active pulsars and magnetars. 

Additional systematic errors in the bare-ACIS data of 1999--2004 were caused 
by the choice of 
Graded telemetry format\footnote{See \S\,6.4.12 in POG.}
 (to avoid telemetry
saturation from the very bright SNR), which made it impossible to correct
the spectra for Charge Transfer Inefficiency (CTI), and an off-axis placement
of CCO (to image the SNR around the chip center), which blurred the CCO image
and hampered the search for a possible 
pulsar
wind nebula). Finally, in all but one of the 1 Ms observations the CCO
was placed at the boundary between two CCD nodes, where ``bad pixels''
contaminate the data.

In the ACIS/HETG observation of the CCO, which did not suffer
from pileup and was telemetered in Faint mode, the dispersed CCO
spectrum was very strongly contaminated by the bright SNR background, while
the zeroth order image had too few counts 
and was contaminated by the dispersed SNR image.
In addition, there were virtually no source counts above 4 keV,
where the ACIS/HETG effective area is very small.

The CCO was also observed with the EPIC instrument onboard the
{\sl XMM-Newton} observatory \citep{mereghetti02}. However,
because of the relatively low angular resolution,
the CCO data were strongly contaminated by bright SNR filaments
in the CCO vicinity.

Thus, despite the very deep observations taken, systematic errors,
particularly those caused by pileup, have precluded an accurate
spectral analysis of the CCO, crucial for understanding its nature.
Therefore, we carried out another {\sl Chandra} ACIS observation of the CCO, 
using a 
subarray of the only activated ACIS-S3 chip.
The much shorter frame time of 0.34 s in this configuration allowed us to obtain
a CCO spectrum virtually undistorted by  pileup, and to search for pulsations
with a period $\gtrsim 0.68$ s. Placing the target very close to the optical
axis, we were able to image the immediate vicinity of the CCO with 
a high spatial resolution and search for compact extended emission (e.g., a pulsar wind
nebula) that might be associated with the CCO.   

In this paper we present the results of this observation.
The data reduction is described in \S2.
In \S3 we present the observational results,
including the image analysis (\S3.1), spectral 
analysis (\S3.2), and timing (\S3.3).
Implications of the results and the possible nature of
the CCO are discussed in \S4.

\section{Observations and Data reduction \label{reduction}}

We observed the Cas A CCO on 2006 October 19 (ObsID 6690) with the {\sl Chandra} ACIS detector for 70.181 ks in Timed Exposure mode,
using Faint telemetry format. 
The target was imaged on the ACIS-S3 chip close to the optical axis (with
a standard $20''$ Y-offset to move the target from the node boundary).
  To minimize the photon pileup and improve time resolution, we used 
a 100 pixel subarray ($8.3'\times 0.82'$ field of view [FOV]) near 
the chip readout 
and turned off the other ACIS chips.  
In this observational setup, the frame time is 0.34104 s, 
which consists of 0.3 s exposure time 
and readout (dead) time of 41.04 ms required to transfer charge 
from the image region to the frame store region.  Therefore, the effective 
target exposure time was 61.735 ks.  There were no periods of substantially 
enhanced background during the observation.

We reduced the data using
{\sl Chandra} Interactive Analysis 
of Observations (CIAO) software\footnote{See 
\url{http://cxc.harvard.edu/ciao/}}, 
ver.\ 4.1.2 (CALDB ver.\ 4.1.2).  
For the timing analysis, we transformed the event times of arrival 
to the solar system barycenter using the {\tt barycen} tool.  In our image 
analysis we used  MARX\footnote{MARX (Model of AXAF Response to X-rays) is a suite of programs designed to enable the user to simulate the on-orbit performance of the {\sl Chandra} satellite.  See \url{http://space.mit.edu/ASC/MARX/}}
 and {\sl Chandra} Ray Tracer (ChaRT) software\footnote{The software is available at \url{http://cxc.harvard.edu/chart/}.}.
We used  XSPEC (ver.\ 12.0.4) for the spectral analysis.

\begin{figure}
\includegraphics[scale=0.45]{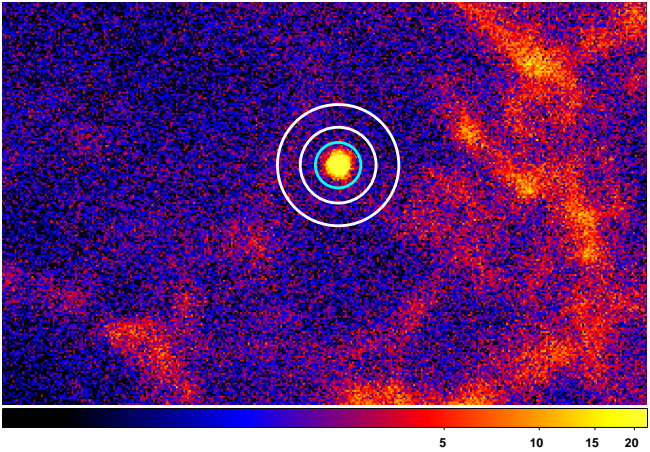}
\caption{{\sl Chandra} ACIS image,
$51''\times 39''$, of the Cas A CCO and its surroundings
in the
0.6--6 keV band, after applying the subpixel event reposition tool
by \citet{mori}.
The pixel size is $0.123''$ (i.e., 1/4 of the original ACIS pixel size).
 The inner circle around the CCO shows the source extraction region of
$1.476''$ radius used for the spectral analysis, while the annulus between
$2.46''$ and $3.94''$ (white circles)
is the background extraction region.  The CCO is surrounded by non-uniform
SNR emission, with the average surface brightness of
89 counts arcsec$^{-2}$ in the $2.46''$--$3.94''$ annulus.
}
\label{imagen}
\end{figure}

\section{Data analysis and results \label{analysis}}

\subsection{Image \label{image}}

The $51''\times 39''$ image of the CCO field is shown in Figure 1.
The CCO looks like a pointlike source embedded in the SNR background.  
Using the CIAO {\tt wavdetect} tool, we found the centroid of the CCO image 
at the coordinates  
$\alpha=23^{\rm h} 23^{\rm m} 27.952^{\rm s}$ and 
$\delta=+58^{\circ} 48' 42.57''$ (J2000).  
The CIAO {\tt celldetect} tool yielded 
$\alpha=23^{\rm h} 23^{\rm m} 27.956^{\rm s}$ and 
$\delta=+58^{\circ} 48' 42.58''$. 
The differences of $0.03''$ and $0.01''$ in right ascension and declination, respectively,
exceed the formal centroiding uncertainties (e.g., $0.006''$ per coordinate 
for the {\tt wavdetect} measurement). They, however, are much smaller than 
the uncertainty of the {\sl Chandra} absolute astrometry,
$\approx 0.2''$ for each of the coordinates at the confidence level of 68\%
\citep{pavlov2009a},
as 
estimated from the empirical distribution of radial offsets of 
X-ray positions with respect to the accurately known celestial locations 
for a sample of point sources\footnote{See \S\,5.4 and Fig.\ 5.4 
in POG.}.
The measured CCO coordinates are consistent with the most accurate of the previously measured coordinates \citep{fesen2006}, within the uncertainties.

\begin{figure}
\includegraphics[scale=0.5]{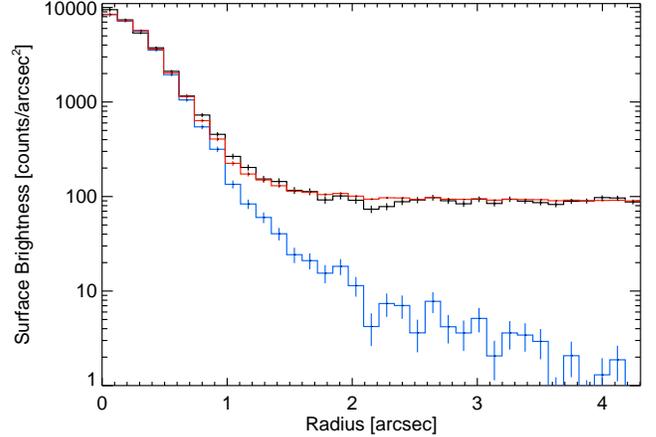}
\caption{Radial profiles of the observed data (black) and the MARX-simulated PSF (blue) in the 0.6--6 keV band. The surface brigthness was measured in the images with pixel size of $0.0615''$ (1/8 of the original ACIS pixel) in 35 circular annuli with $0.123''$ widths.  The simulated PSF profile is for the MARX DitherBlur parameter of $0.30''$.
The red histogram shows the sum of the sumulated PSF and the model constant background, 89.6 counts arcsec$^{-2}$.}
\label{fig:psf_marx}
\end{figure}
 
To look for an extended component in the CCO image,
 such as a pulsar wind nebula (PWN), and 
to choose an optimal extraction aperture for the spectral and timing analysis, 
we simulated a point source observation using ChaRT and
MARX and compared the results with the observed count distribution. 
To reach a subpixel spatial 
resolution, we 
applied the subpixel event reposition 
method 
\citep{mori}
to both the observed and simulated images.
For the simulation, we chose the spectral model {\em wabs $\times$ nsa} 
with the 
best-fit parameters (see \S\ref{spectral} and Table \ref{table:spectra}) 
and simulated an ACIS-S3
observation for the same position on the detector and the same
 exposure time, 61.735 ks, as those of the actual observation.
The width of the simulated point spread function (PSF) depends on the 
value of the MARX parameter DitherBlur. This parameter 
accounts for the ACIS pixelization and aspect
reconstruction errors, which may be different for different observations. 
We simulated the PSF for a number  
of DitherBlur values, from $0.20''$ to $0.40''$,
and found that the best match of the simulated PSF to the core of the
observed image is provided by DitherBlur $\simeq 0.30''$. 
Figure \ref{fig:psf_marx} shows the simulated PSF radial profile 
and the radial distribution of the detected events up to $4.3''$ from the source
centroid.
To calculate and plot these profiles, we rebinned the observed and simulated
images to 1/8 of the original ACIS pixel and measured the numbers of counts
in circular annuli with widths of $0.123''$ (1/4 of the original pixel
size). We see from this figure that the 
observed
radial profile exceeds the simulated PSF at $r\gtrsim 1''$, 
remaining approximately constant at $r\gtrsim 2''$,
with the 
average surface brightness
of $89.6\pm 1.5$ counts arcsec$^{-2}$ in the $2''<r<4''$ annulus. 
The morphology of this extended emission on larger scales (see Fig.\ 1)
suggests that it originates from the Cas A SNR (possibly belongs to a faint
SNR filament), i.e., the extended emission 
is not a nebula generated by the CCO. This conclusion is 
supported by the presence of emission lines in the spectrum of
 this extended emission
(see \S\,3.2), which are not expected in the synchrotron spectrum of a PWN. 

The sum of the simulated PSF and the uniform surface brightness of
89.6 counts arcsec$^{-2}$,
shown by the red histogram
in Fig.\ \ref{fig:psf_marx},
is generally very 
close to the observed radial profile within the $r<4.3''$ circle. 
It lies slightly below the observed data (i.e., the observed profile is
slightly broader than the simulated one) at $0.7''\lesssim r \lesssim 1.2''$,
but the statistical significance of this difference is marginal
(e.g., the largest discrepancy, in the 7th annulus, is significant at
the $2\sigma$ level).
The difference might be caused by a nonuniformity of the extended
emission component at such radii
or a minor inaccuracy of the MARX simulation, but it is hard to believe
that the data excess is due to a PWN-like emission in this narrow region.
An additional support for the pointlike structure of the CCO image
is provided by the image deconvolution with the aid of the {\tt arestore}
script in CIAO (based on the Lucy-Richardson algorithm), performed by \citet{kargaltsev09}
for the comparison with PSR J1617--5055 that
{\em is} embedded in a PWN. These authors conclude that the deconvolved
image of the Cas A CCO ``preserves the pointlike appearance with no
extended structure'' (see Fig.\ 4 of that paper).
Thus, we conclude that our observation does not provide any substantial
evidence for a PWN around the Cas A CCO.

\begin{figure}[ht]
\hspace*{-0.7cm}
\includegraphics[scale=0.55,angle=0]{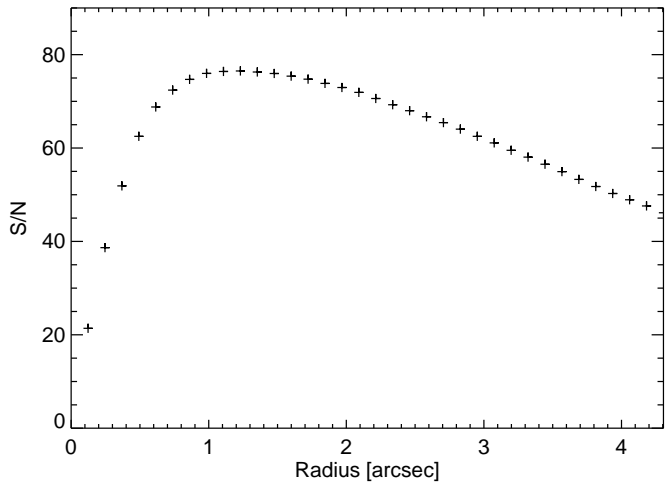}
\caption{Signal-to-noise ratio as a function of circular aperture radius.}
\end{figure}

\vspace{3cm}
\subsection{Spectral analysis \label{spectral}}

As the CCO is embedded in the 
significant SNR background, the 
source extraction
region should be smaller than that for an isolated point source. 
To find an optimal extraction radius, we modeled the dependence of the
signal-to-noise ratio ($S/N$)
on the radius $r$ of the source extraction aperture 
in the 0.6--6 keV band (Figure 3),
using the above-described MARX simulation and the measured background
surface brightness in the CCO vicinity.
The $S/N$ shows a flat maximum in the range $1.0''\lesssim r \lesssim
1.5''$, with the peak value $S/N= 76.5$ at $r=1.23''$. To minimize the
spectral distortion caused by
the increase of the PSF size with growing energy (i.e., the spectrum extracted
from the PSF core is somewhat softer than the true spectrum), we chose
the source aperture radius of $1.476''$ (three ACIS pixels), near the
larger radius of the flat maximum.
The MARX simulation shows that
this circle contains 
95.4\% of total point source counts
and 
94.8\% of total energy flux.
Because the SNR background is not uniform (see Fig.\ 1), we have to
choose the background extraction region sufficiently close to the CCO but not
too close, in order
 to prevent the background ``contamination'' by the wings of the
CCO's PSF.
After several trials,
we chose the $2.46'' < r < 3.94''$ (5--8 ACIS pixels)
 annulus as the background region\footnote{To 
examine the effect of the source and background
aperture sizes on the results of spectral fitting,
we also 
fit the spectra extracted from the source apertures of $1.0''$ and $2.0''$ radii,
and used a few other annular background extraction regions.
The spectral 
parameters obtained from those fits are generally consistent with those
reported below (the differences are within the $2\sigma$ uncertainties).}. 

We used the CIAO {\tt psextract} script to extract the spectra.  
The $1.476''$ radius aperture 
contains 
7016 counts in the 0.6--6.0 keV range, 
of which 
about 607 counts belong to the background.
This corresponds
to the source count rate of 
$0.104\pm 0.02$ counts s$^{-1}$ in the source
extraction region (or 
$0.109\pm 0.02\,\, {\rm counts}\, {\rm s}^{-1}$ 
after correcting
for the finite size of the extraction aperture).
The source count rate of $\simeq 0.033$ counts per frame
corresponds to the pile-up fraction (i.e., the ratio of the number
of frames with two or more events to the number of frames with one or more
events) of about 1.6\%, a factor of 10--15 lower than in the previous ACIS 
observations with the frame time of 3.24 s.

We grouped the extracted spectrum with a minimum of 
30 counts per spectral bin 
and fit 
it with several spectral models.
The results are shown in Figures 4 and 5, and Table 1.
The fit with the absorbed power-law (PL) model 
yields the photon index $\Gamma\approx 5$, 
much larger than $\Gamma =1$--2 for the X-ray spectra of active pulsars, 
and larger than  $\Gamma=3$--4 obtained from the previous observations 
in which the CCO spectrum was considerably piled up.  
The observed flux, 
$F\approx 
6.7\times 10^{-13}$ erg cm$^{-2}$ s$^{-1}$ in the 0.6--6 keV band 
($7.0\times 10^{-13}$ erg cm$^{-2}$ s$^{-1}$ after correcting for
 the encircled energy fraction),
is slightly lower than that found in the previous observations.
The fit is 
marginally acceptable in terms of minimum $\chi^2$ 
($\chi_\nu^2=1.21$ for $\nu=125$ dof),
but it shows excess data counts at lower energies ($\lesssim 1.2$ keV) 
and a deficit of data counts at higher energies ($\gtrsim 4.5$ keV). 
This suggests that the true spectral shape is different from the PL.  
In addition, 
the  PL fit gives a rather large value for the hydrogen 
column density, $\nh \approx 2.8$, 
 which not only strongly exceeds the  total Galactic HI column density
in that direction 
($N_{\rm HI,22}=0.4$--0.7; see \citealp{dickey,kalberla})
but is also substantially greater than 
typical values $\nh \approx 1.2$--1.3 ($\nh = 1.5$ for a maximum value)
in the CCO vicinity found from the fits of the SNR spectra with the absorbed
two-component plasma emission model in the 1 Ms exposure
data \citep{yang}. 

The fact that the observed spectrum is softer at higher energies than 
the best-fit PL model (even with the large $\Gamma$) suggests that a 
thermal model could provide a better description.  
The fit with the absorbed blackbody (BB) model 
gives\footnote{ We add the superscript $^\infty$ to the temperature and radius to show that these parameter are quoted as observed by a distant observer, without
any corrections for the gravitational effects.
The temperature, radius and bolometric luminosity
 ``at infinity'' are connected with those as observed at the NS surface 
as follows: $T^\infty=Tg_r$, $R^\infty=R/g_r$, and $L_{\rm bol}^\infty
=L_{\rm bol}g_r^2$,
 where $g_r=(1-2GM/Rc^2)^{1/2}$ is the gravitational redshift parameter.}
 temperatures $kT^\infty=0.39$--0.41 keV,
hydrogen column densities $\nh=1.2$--1.4, 
radii of equivalent emitting sphere 
$R^\infty \approx (0.8$--$1.0)D_{3.4}$ km
($R^\infty/D\approx 
0.27$ km/kpc), and bolometric luminosities
$L_{\rm bol}^\infty \sim 3\times 10^{33}$ erg s$^{-1}$.
The temperatures are lower, and the radii are larger than those obtained 
from the piled-up spectra from the previous observations. 
The quality of the BB fit is considerably lower than that of the PL fit.  
Not only the BB fit  gives a larger $\chi_\nu^2$ 
(1.54 vs.\ 1.21),
but it
shows a strong excess of data counts at $E \gtrsim 4$ keV.  
Therefore, we conclude that the BB model gives a poorer description 
of the CCO spectrum than the PL model.  

\begin{figure*}[ht!]
\begin{center}
\includegraphics[scale=0.65,angle=-90]{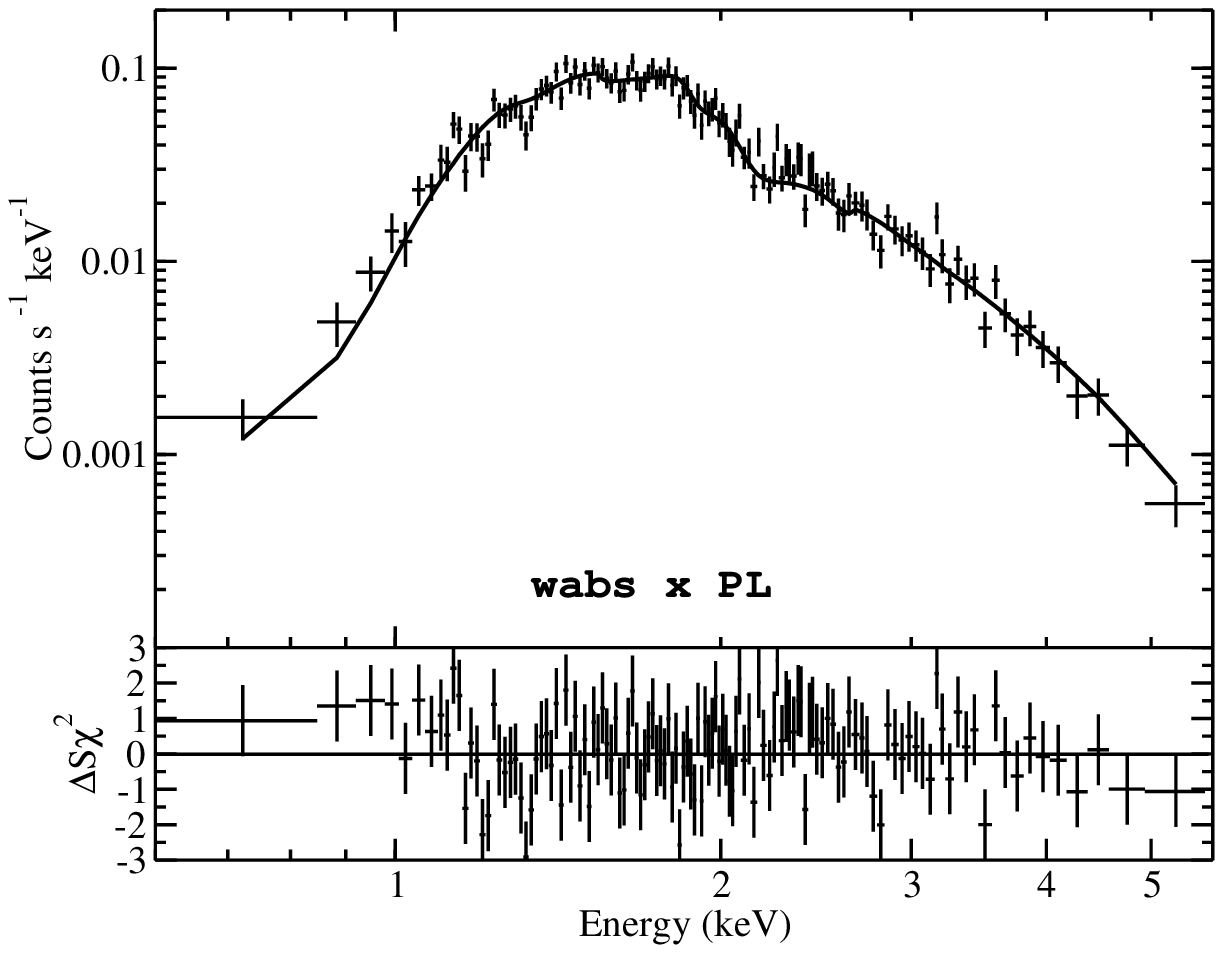}
\includegraphics[scale=0.65,angle=-90]{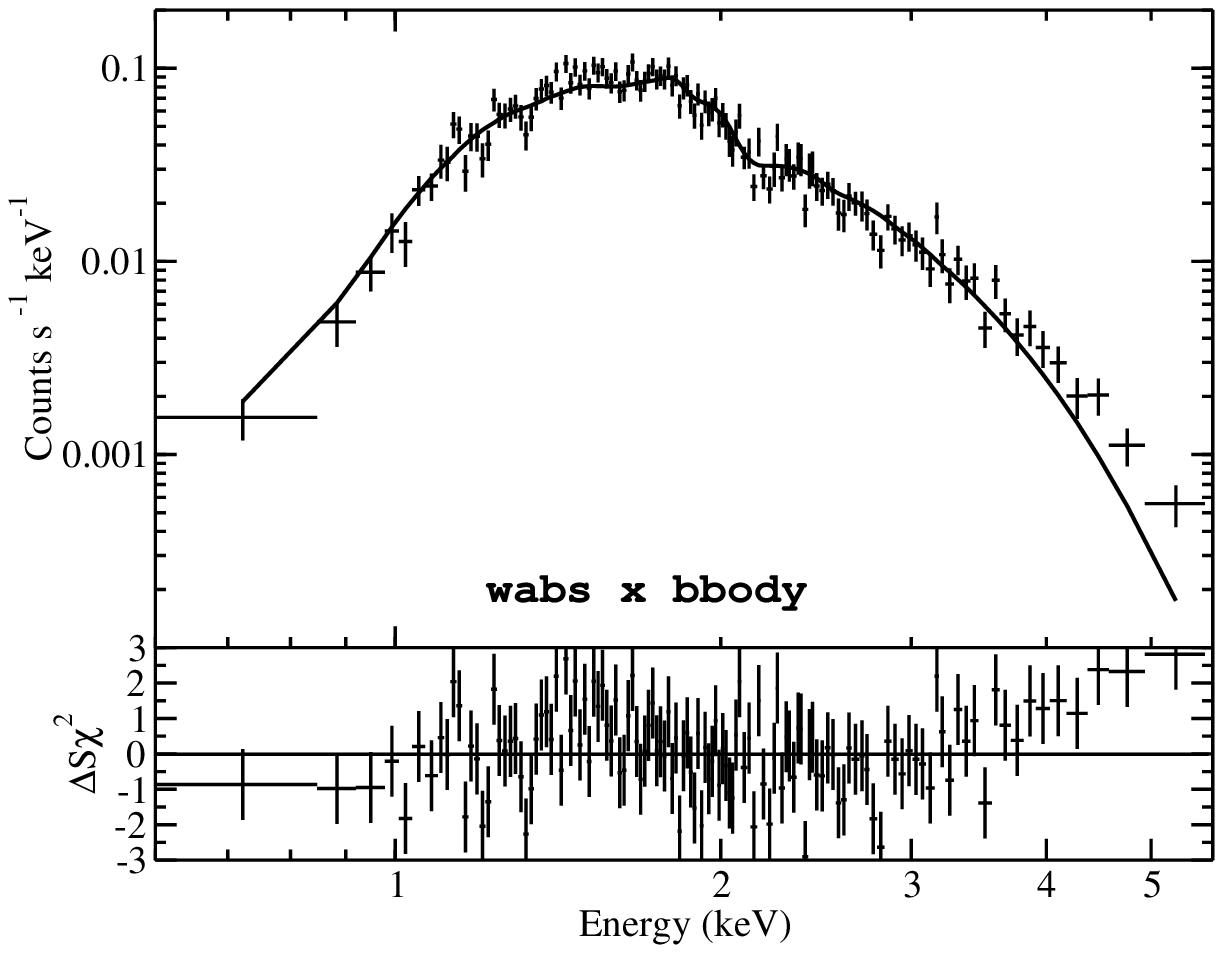}
\includegraphics[scale=0.65,angle=-90]{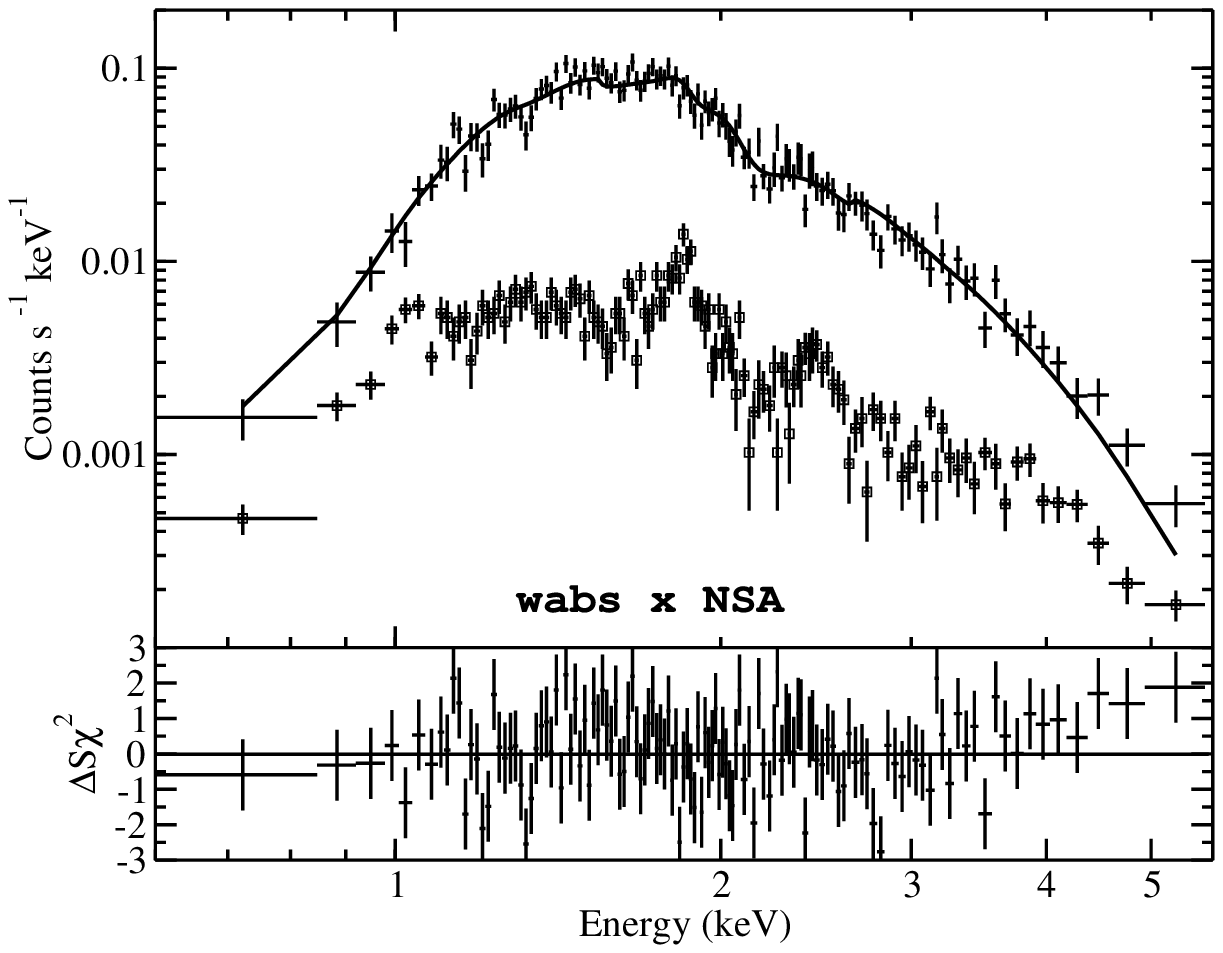}
\includegraphics[scale=0.65,angle=-90]{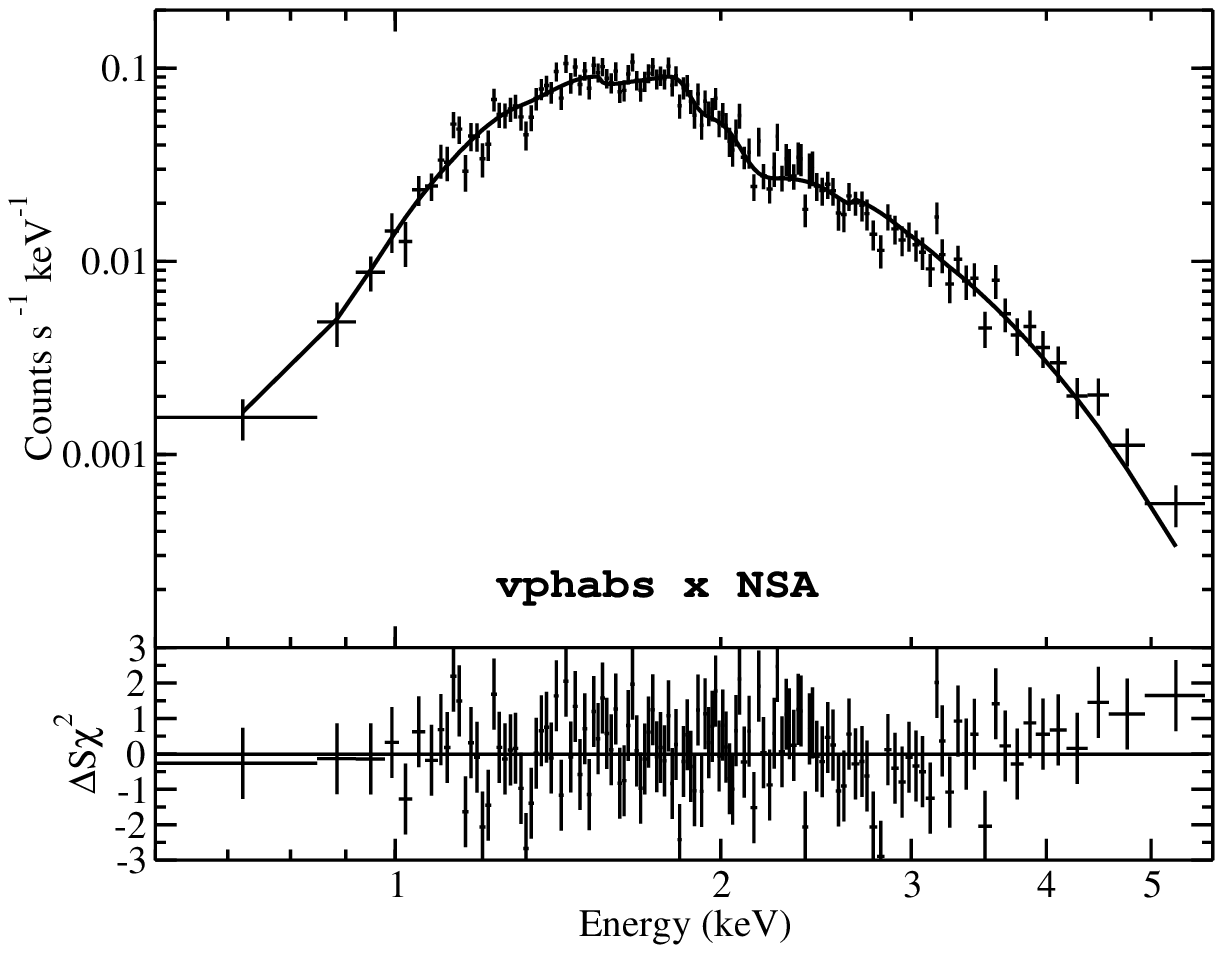}
\includegraphics[scale=0.65,angle=-90]{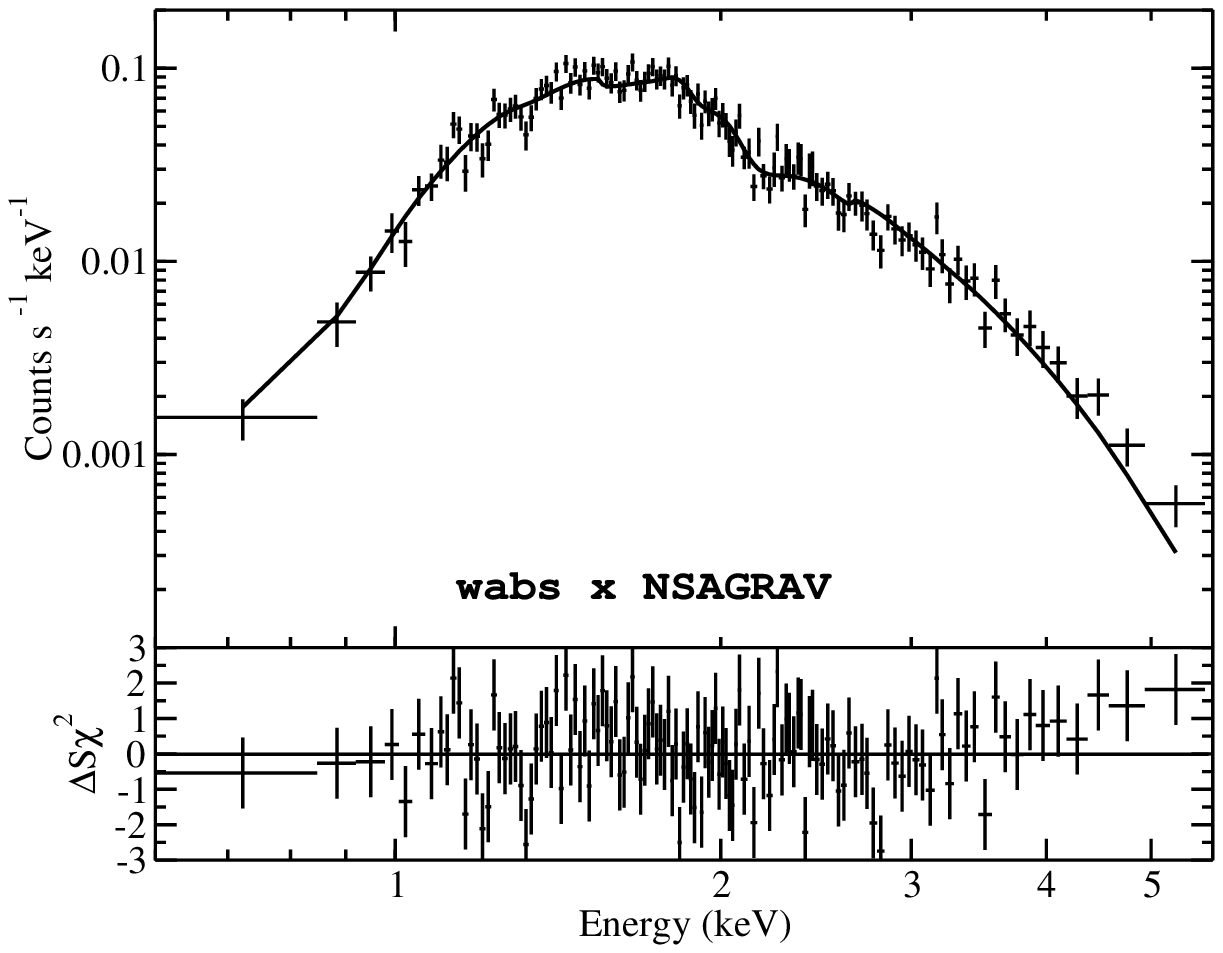}
\includegraphics[scale=0.65,angle=-90]{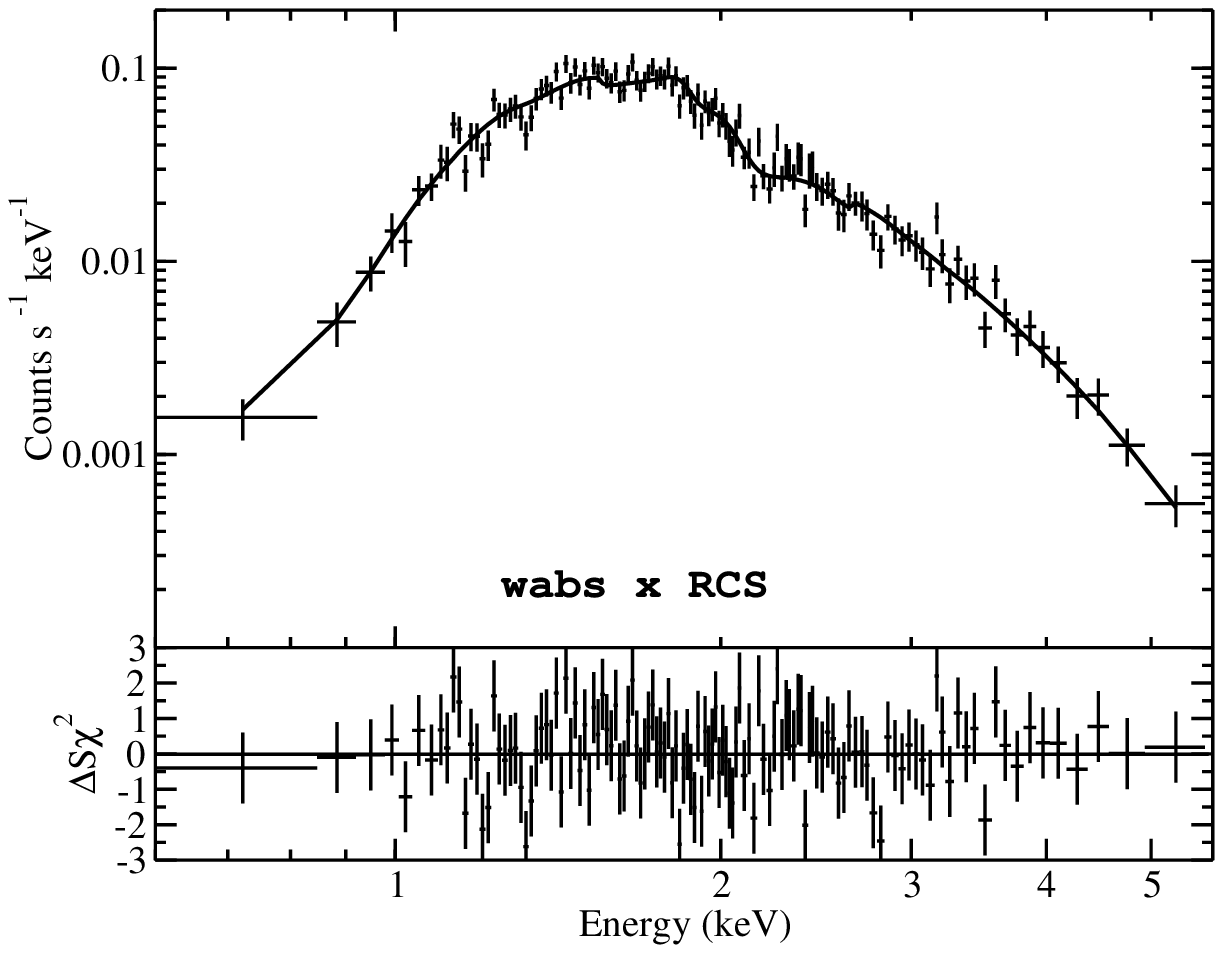}
\caption{Fits of the CCO spectrum extracted from the $1.476''$ aperture with
absorbed one-component spectral models (see Table 1 for the fitting parameters).The {\em wabs}$\times${\em NSA} panel also shows the background contribution. }
\end{center}
\label{fig:fits1}
\end{figure*}

As the observed spectrum is harder than a BB but softer than a PL 
at higher energies, similar to the spectra of hydrogen (or helium) 
NS atmosphere (NSA)\footnote{The hydrogen NSA models are  dubbed {\em nsa} 
and {\em nsagrav} in XSPEC.  The {\em nsa} models are for fixed NS mass 
and radius, $M=1.4 M_\odot$ and $R=10$ km ($R^\infty = 13.06$ km, gravitational
acceleration at the NS surface $g=2.43\times 10^{14}$ cm s$^{-2}$); 
in these models the distance is the fitting parameter.  
The {\em nsagrav} models include a set of NSA models on a mass-radius grid; 
therefore, they allow one to fit the mass and radius for a given
distance.} models \citep{pavlov1995,zavlin1996}, we fit it with 
a number of such models. An example of such a fit, for  a hydrogen 
atmosphere model with a 
relatively low magnetic field 
($B< 10^{10}$ G)
and  NS mass $M=1.4 M_\odot$ and radius $R=10$ km ($g_r=0.766$), is 
shown in Figure 4.
In terms of $\chi_\nu^2$,
the NSA fits are considerably better than the BB fit.
Most of the NSA fits show 
an excess of data counts 
at high energies
 (e.g., $18.3\pm 7.7$ counts in the 4.5--6 keV for the
{\em wabs$\times$nsa} fit shown in Figure 4), 
but it is not so strong as in the BB fits. 
Such an 
excess might be caused by the 
(small) pileup effect 
(which hardens the count spectrum, as mentioned in \S\ref{intro}),
 but our estimates show that only a few 
such events are expected in our observation.
The NSA models with high magnetic fields ($B
> {\rm a\,\, few}\,\, 
10^{12}$ G) give 
worse fits than the low-field NSA models 
(in particular, a stronger excess at higher energies) because the
spectra of high-field NSA models are closer to BB spectra \citep{pavlov1995}.  
In comparison with the BB model, the low-field NSA models give a factor of 
2 lower effective temperatures, $kT_{\rm eff}^\infty=0.17$--0.20
 keV,
and a factor of $\sim 7$ larger
radius-to-distance ratio, $R^\infty/D = 1.3$--1.7 km/kpc,
for $M=1.4 M_\odot$ and $R=10$ km.
For this mass and radius, the best-fit distances, $D=7.5$--10.1 kpc,
substantially exceed the measured distance to Cas A, 
suggesting a smaller size of the emitting region\footnote{ Although the spectral shape
depends on the assumed $M$ and $R$, in contrast with the BB spectrum, the
dependence is weak, so that the radius can be rescaled as $R\propto D$
for crude estimates.}, $R^\infty\sim 
4$--6 km, 
considerably larger than for the BB fit but 
still too small for a NS radius [and comparable to the Schwarzschild 
radius, $R_s=2.953 (M/M_\odot)$ km].
This suggests that either only a fraction of the NS surface is heated
up to X-ray temperatures
(in which case we should expect pulsations in the X-ray emission)
 or the CCO is not a NS at all (e.g., a strange
quark star [SQS], whose radius 
can be considerably smaller than that
of a NS). To explore the latter possibility, we 
fit the CCO spectrum
with 
the {\em nsagrav} models, which allow one to vary $M$ and $R$.
An example of such a fit, in which the  mass was fixed
at $M=0.25 M_\odot$, the distance fixed at $D=3.4$ kpc,
and the radius was varied,
is presented in Figure 4, for the best-fit parameters
$R=5.077$ km, $T_{\rm eff}= 198$ eV, and $\nh=1.57$. The fit
residuals are virtually indiscernible from those of the {\em nsa} fit
(including the slight data excess at higher energies),
and the $T_{\rm eff}^\infty$ (but not $T_{\rm eff}$)
 and $\nh$ values are also almost the same.
The corresponding bolometric luminosity is 
$L_{\rm bol}^\infty \approx 4.2\times 10^{33}$ erg s$^{-1}$, of which 91\% is
emitted in the 0.6--6 keV band.
 This fit confirms that 
the observed CCO emission might be interpreted as radiation emergent
from a light-element atmosphere of an object more compact than a NS.
We will discuss such interpretations in more detail in \S4. 

\begin{figure}
 \includegraphics[scale=0.65,angle=-90]{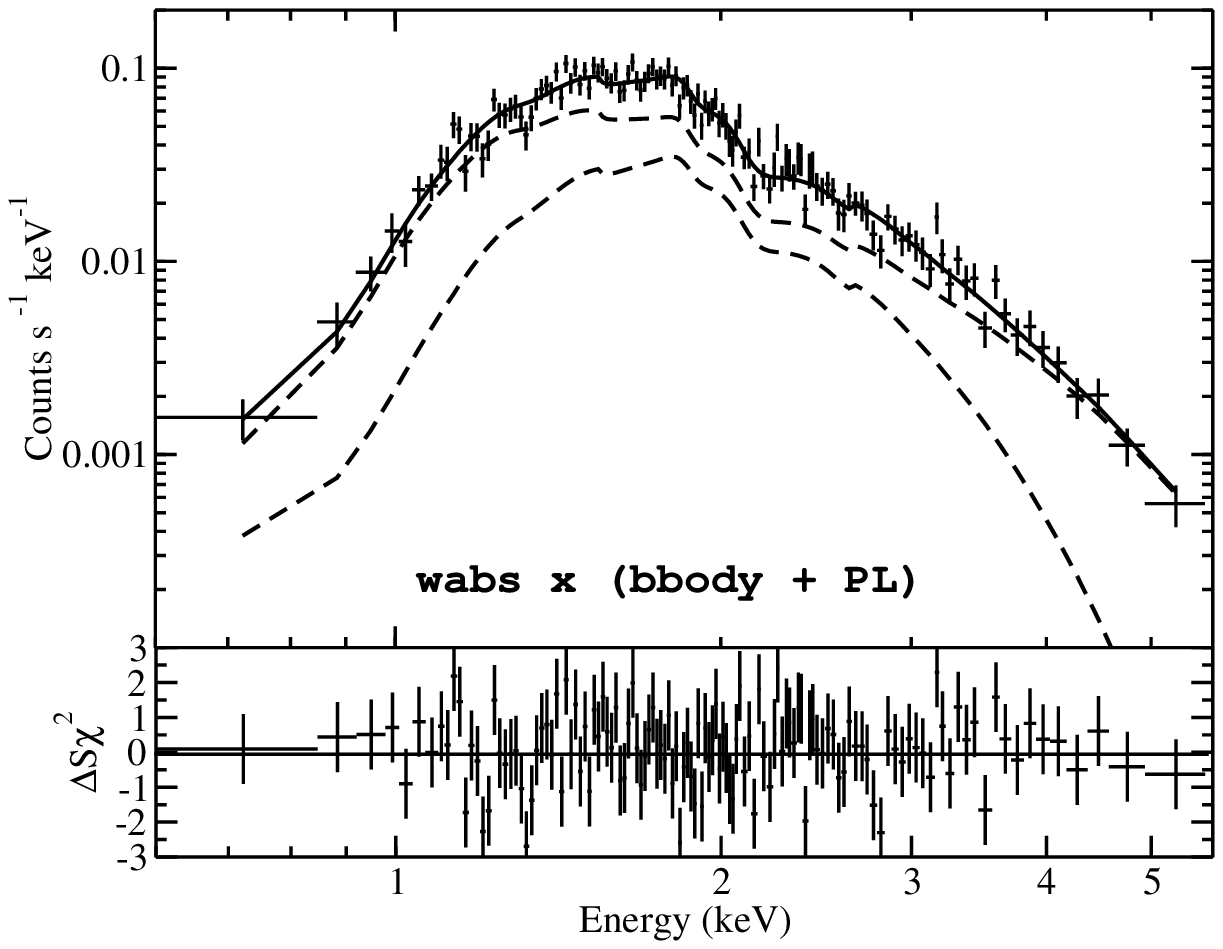}\\
 \includegraphics[scale=0.65,angle=-90]{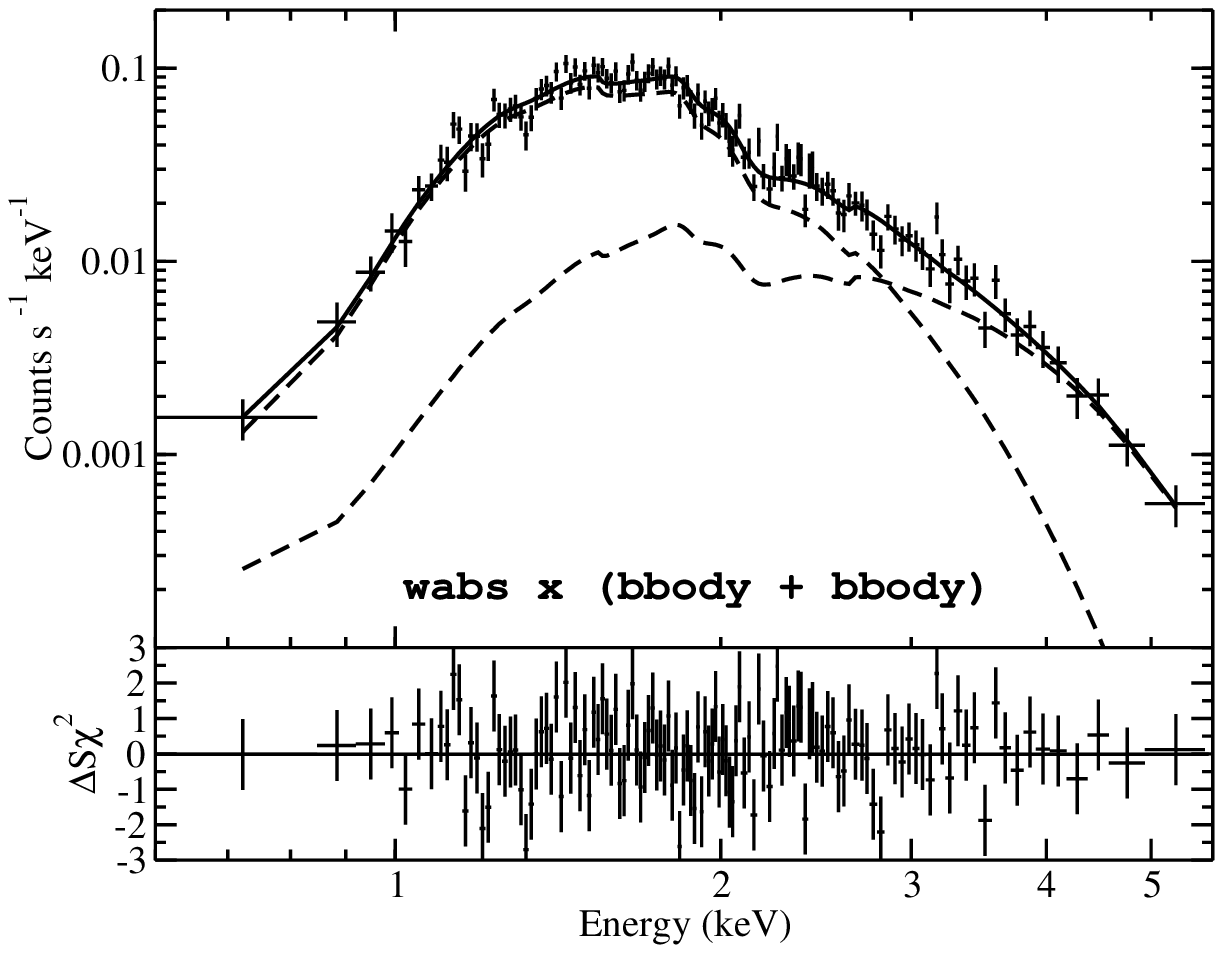}\\
 \includegraphics[scale=0.65,angle=-90]{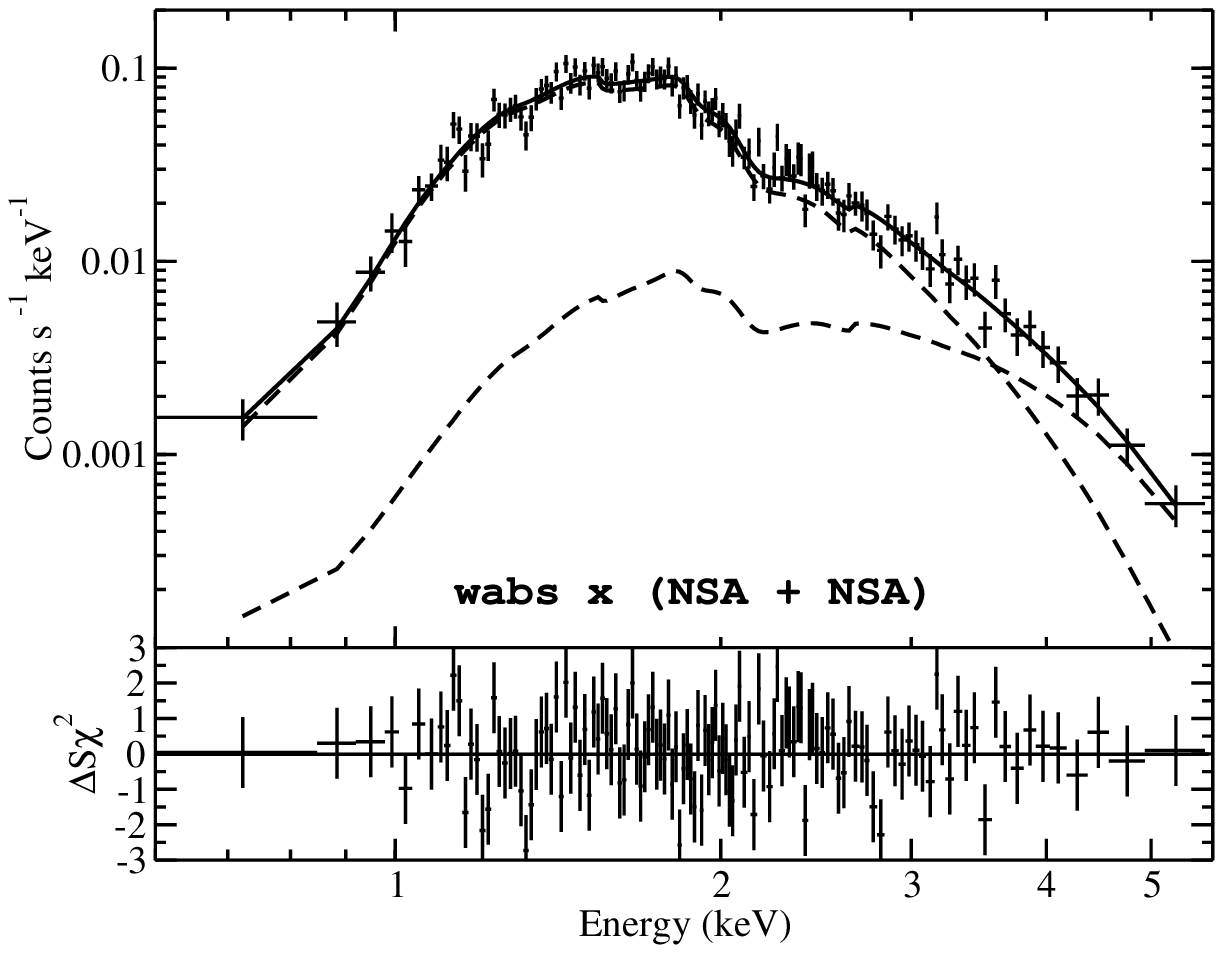}
\caption{Fits of the CCO spectrum extracted from the $1.476''$
aperture with absorbed two-component models (see Table 1 for the fitting parameters). The component contributions are shown by dashed lines.}
\label{fig:fits2}
\end{figure}

If the CCO has a strong magnetic field and its magnetosphere contains
 a large number of energetic electrons, these electrons can comptonize
 the NS thermal emission
and generate a high-energy tail in the CCO spectrum.
This effect can be crudely accounted for by the
``resonance Compton scattering'' (RCS) XSPEC model
\citep{lyutikov,rea}. 
The fit of the CCO spectrum with the RCS model is statistically good
($\chi_\nu^2 = 1.11$), but it yields very broad ranges for the 
temperature 
of the seed BB emission, $kT\approx 0.1$--0.2 keV, and the normalization,
$\approx 3\times 10^{-5}$--$10^{-2}$, which are strongly anti-correlated
(i.e., lower temperatures correspond to larger normalization parameters).
The fit gives
    the 
resonant scattering optical depth $\tau_{\rm res}=1$ 
(at the lower boundary of the interval $\tau_{\rm res} = 1$--10 included in the
XSPEC set of the RCS models),
and the characteristic velocity (in units of $c$)
of the magnetospheric electrons $\beta_T = 0.24$--0.35.
We see that the seed BB temperature is considerably lower than that
found from the BB fit,
which hints that the size of the
emitting region is larger
 than 
BB radius, but, unfortunately,
it remains unclear how the size could be estimated from the model
normalization (N.\ Rea, priv.\ comm.). We should note, however, that,
since 
the RCS model itself is 
substantially oversimplified (e.g., it assumes a one-dimensional
motion of non-relativistic
electrons with a rectangular velocity distribution along the
radial direction), 
the inferred fitting parameters
may be quite different from
the actual physical parameters. Moreover, there is no even circumstantial
evidence that the Cas A CCO has 
a magnetosphere densely populated by energetic particles
 (such as the nonthermal X-ray spectra in rotation-powered
pulsars or the  
hard tails  
seen in magnetars even in quiescence). Therefore,
the good fit may simply mean that the RCS model just 
mimics the observed CCO spectrum whose true origin may be quite different.

The quality of the spectral fits can be improved if we use various 
two-component models. Examples of such fits are shown in Figure 5 and Table 1.
For instance, the PL+BB model (which is commonly used for the description
of the X-ray spectra of rotation-powered pulsars, AXPs and SGRs) provides a
good fit ($\chi_\nu^2 = 
1.12$; 
no excesses or deficits at lower and higher
ends of the spectrum), but the slope of the PL component is still very
steep ($\Gamma\approx 4.6$), the size of the BB-emitting region is still
much smaller than the NS radius ($\approx 1.2$ km at $D=3.4$ kpc),
and the hydrogen column density is too large ($\nh \approx 
2.3$).

The BB+BB model (which can be considered as a crude model
for thermal emission from the surface with a non-uniform temperature)
gives a good fit ($\chi_\nu^2=
1.11$), with 
$kT_{\rm soft}^\infty\approx 0.3$ keV, 
$R_{\rm soft}^\infty\approx 2$ km,
$L_{\rm bol,soft}^\infty \approx 4.1\times 10^{33}$ erg s$^{-1}$
 for the low-temperature (soft) component,
and $kT_{\rm hard}^\infty\approx 0.6$ keV, 
$R_{\rm hard}^\infty\approx 0.25$ km,
$L_{\rm bol,hard}^\infty \approx 0.6\times 10^{33}$ erg s$^{-1}$
for the high-temperature (hard) component (the radii and luminosities
are for $D=3.4$ kpc). Although the addition of another BB component 
has resulted in a larger size, this size is still much smaller than the NS radius.

Obviously, one could obtain a larger emitting size using a two-component
NSA model. The NSA+NSA fit with low-field {\em nsa} models
gives the temperatures $T_{\rm eff,soft}^\infty\approx 0.14$ keV and
$T_{\rm eff,hard}^\infty\approx 0.4$ keV. 
The range of distances for the 
soft component, $D=1.8$--5.9 kpc, includes the range of 3.3--3.7 kpc
inferred by \citet{reed}
for the Cas A SNR.
Moreover, 
if we rescale $R_{\rm soft}^\infty =13.06$ km to $R_{\rm soft}^\infty\approx 11$ km, still consistent with a NS radius, it would correspond to the best-fit
distance of 3.4 kpc, equal to the most probable distance to Cas A.
The similarly scaled size for the hard component
is 
$\sim 0.4$ km, for $D=3.4$ kpc.
For these effective temperatures and scaled radii, the bolometric
luminosities are 
$L_{\rm bol,soft}^\infty \approx 5.7\times 10^{33}$ erg s$^{-1}$ and
$L_{\rm bol,hard}^\infty \approx 0.4\times 10^{33}$ erg s$^{-1}$.
 Thus, we cannot rule out
that the CCO is a NS with a nonuniform 
surface temperature, although it is hard
to explain the nonuniformity without assuming a strong magnetic field
of a special topology (see \S4).

In all the fits described above, we see some structure in the residuals, 
in the range of $\approx 1$--3 keV, which hints 
at the presence of spectral lines or photoabsorption edges. 
If real, such spectral features could be either intrinsic features
in the CCO spectrum (e.g., 
similar to the absorption lines detected
in the spectrum of the CCO 1E\,1207.4--5209; 
\citealp{sanwal}) 
 or they might be caused
by photoabsorption between the source and the observer 
if the element abundances are different from those adopted 
in the {\em wabs} photoabsorption model used in the above fits. 
To explore the latter possibility, we applied the {\em vphabs} 
photoabsorption model with variable abundances
together with the NSA model.
Fitting the abundances of the 
elements with photoionization 
edges in the 0.6--3 keV range (Fe, Ne, Mg, Si, S), we found that 
the Fe, Ne, S and Mg 
abundances were poorly constrained and consistent with the solar values, 
while the Si abundance showed 
some excess\footnote{Possible additional 
absorption at the Si-K photoionization edge, 1.84 keV, has been noticed by
\citet{stage} 
in previously observed spectra of the CCO. These
authors interpret it as caused by material in the Cas A SNR.}. 
Fixing the abundances of
all the elements but Si at their solar values, we obtained the Si abundance
of $3.2\pm 1.3$, without substantial changes of the other fitting parameters
(see Table 1). 
The fit shows a marginal improvment with respect to the {\em wabs$\times$nsa}
fit 
($\chi_\nu^2=1.19$ vs.\ 1.24) and somewhat smoother residuals in
the 1.5--2.5 keV range, but it does not affect the structure at 1.1--1.4 keV
and the apparent absorption feature around 2.8 keV.

We have checked that such features could not be caused by inaccurate
subtraction of the SNR background, which does not show spectral lines
at these energies (see the panel {\em wabs$\times$nsa} in Figure 4).
However, 
our simulations of ACIS count rate spectra for continuum spectral models
show that similar ``features'' often appear 
because of statistical fluctuations. We also examined the spectra
obtained in the previous, much longer observations of the CCO
 and found no features
at these energies for most of them\footnote{We should mention, however,
that the 
stronger parallel CTI effect in those observations might wash out faint
spectral features.}.
Therefore, we 
conclude that, most likely, our observation does not show
real spectral features in the CCO spectrum.

\begin{figure}
\includegraphics[angle=-90,scale=0.85]{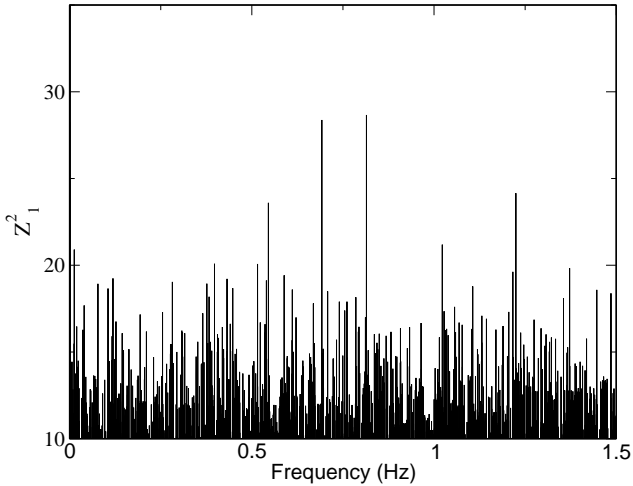}
\caption{Power spectrum $Z_1^2(f)$ for the Cas A CCO.}
\label{powerspec_1.0_z1}
\end{figure}

\subsection{Timing \label{timing}}

To search for the CCO period, we used the
arrival times for the 
6756 events extracted from the $1.23''$ radius aperture 
(which provides maximum $S/N$; see Figure 3)
in the 0.6--6.0 keV band (the estimated background contribution
is 
421 events).  With the 
frame time $t_{\rm frame}=0.34104$ s, we can search for periods $P>2t_{\rm frame} = 0.68208$ s (frequencies $f<1.466$ Hz).
We calculated the $Z_{1}^{2}$ (Rayleigh) statistic \citep{buccheri1983}
as a function of frequency 
(i.e., the power spectrum) in the range 
$30\,\mu{\rm Hz}< f < 1.466\,{\rm Hz}$
with the step $\Delta f=1\,\mu{\rm Hz}\approx 0.07 T_{\rm span}^{-1}$, where
$T_{\rm span}=70,181$ s is the time span of our observation. The power spectrum
is shown in Figure \ref{powerspec_1.0_z1}. 
The highest peaks in the power spectrum,
$Z_{1}^{2}=
28.66$ and 28.38, are found at the frequencies
 $f=
0.814487$ 
and 0.692145 Hz,
respectively (the frequency uncertainty is about $2\,\mu$Hz for each of the peaks).  
If, for instance, the highest peak were due to the actual pulsations,
it would correspond to the pulsed fraction
$p = p_{\rm obs} (N_S+N_B)N_S^{-1} \approx [2Z_1^2 (N_S+N_B)]^{1/2}N_S^{-1} \pm [2(N_S+N_B)]^{1/2}N_S^{-1}
= 0.098 \pm 0.018$ (for a nearly sinusoidal signal),
where
$N_S=6333$ and $N_B=421$ are
the numbers of source and background counts in the source aperture,
and $p_{\rm obs}$ is the pulsed fraction uncorrected for
the background contribution.
However, the probability of obtaining such a peak by chance
in a noise spectrum
is $f_{\rm max}T_{\rm span} \exp(-Z_1^2/2) = 0.061$,
 which corresponds to
the confidence level of
only $1.87 \sigma$.
Given the low confidence level and the presence of a few other peaks of
similar heights (see Figure \ref{powerspec_1.0_z1}), we have to conclude
that this observation does not show pulsations with a period
$P\gtrsim 0.68$ s.
Following the approach by \citet{groth75},
the upper limits on the pulsed fraction $p$
can be estimated as 13\%, 14\% or 16\% at the confidence levels of
95\%, 99\% or 99.9\%, respectively.

From their analysis of the {\sl XMM-Newton} observation of the Cas A CCO,
\citet{mereghetti02} report somewhat lower values of the pulsed fraction
upper limits (e.g., $p<13\%$ in the 0.4--2 Hz band at the $3 \sigma$ level).
That observation, however,
suffered from the strong, nonuniform
SNR background, whose contribution into
the source aperture could hardly be measured with a
sufficient accuracy.
Indeed, the fact that the CCO's flux reported by
Mereghetti et al.\
is a factor of 3 higher than the value we measured with {\sl Chandra}
(see Table 1)
suggests that the background contribution, $N_B/(N_S+N_B)$,
is 92\% rather than 75\% estimated in that paper,
and the background-corrected
upper limit on the pulsed fraction
is also a factor of 3 higher
than reported by Mereghetti et al.

\section{Discussion 
\label{discussion}}

Our observation of the Cas A CCO with the {\sl Chandra} ACIS detector
in subarray mode has allowed us to obtain the most accurate X-ray spectrum,
virtually undistorted by pileup, 
to look for extended emission in the high resolution image of this source,
and to search for periodicity with periods greater than 0.68 s. 
The results of this observation can be used not only to measure the 
source properties with high precision, but also to put new constraints
on its nature.

The very high X-ray-to-optical flux ratio (e.g., $F_X/F_H > 10^4$; \citealp{fesen2006})
 proves unequivocally that the Cas A CCO is indeed a compact object
associated with the Cas A SNR, not a usual star or a background AGN.
It is generally believed that a compact source born in a 
Type II SN explosion
can be either a NS or a black hole. If the Cas A CCO were a black hole,
its X-ray emission could be only due to accretion of the ambient matter, 
in which case we would expect a substantial variability (as accretion is
an inherently unstable process) and strong emission lines generated
in the hot accreting matter. However,
 the CCO's flux has not shown stastically
significant changes in numerous {\sl Chandra} observations \citep{teter2004},
its X-ray spectrum does not show strong emission lines (\S3.2),
and its X-ray luminosity is too high to be explained by accretion
of SNR material onto the
apparently fast-moving CCO \citep{pavlov2000}.
 Therefore, we conclude that the
black hole interpretation can be ruled out.

Most of the known NSs are rotation-powered pulsars ($\sim 1800$ such pulsars
have been detected in the radio, $\sim 80$ in X-rays, 
$\sim 30$ in $\gamma$-rays,
and $\sim 10$ in the optical).
X-ray observations of rotation-powered pulsars usually show hard 
PL components in their spectra (with photon index $\Gamma\sim 1$--2), 
interpreted as magnetospheric emission, with sharp pulsations. Moreover,
virtually all young, powerful pulsars 
are accompanied by bright PWNe
generated by the winds of ultrarelativistic particles ejected from the
pulsar magnetosphere (e.g., \citealp{kargaltsev08}
). As the X-ray
spectrum of the radio- and $\gamma$-ray-quiet CCO
is very soft 
(almost certainly thermal; see \S3.2), 
and, more importantly, there
are no traces of a PWN around the CCO (\S3.1), 
we have to conclude that it is not
a rotation-powered pulsar.

The usual 
(rotation-powered) pulsar activity could be quenched in a young NS 
if its rotation is too slow or its magnetic field is too weak
to generate strong electric fields
at and above the NS surface, needed for normal pulsar operation.
Very high magnetic fields ($\gtrsim {\rm a\,\, few}\,\, 10^{13}$ G) 
can also inhibit
the pulsar activity (perhaps, by suppressing electron-positron pair
production because of photon splitting in a superstrong magnetic field; \citealp{baring98}),
as we know from observations of magnetars.
Thus, 
 the Cas A CCO could belong to one of the two known types
of young NSs without pulsar activity --- magnetars,
which are characterized by superstrong magnetic fields
($B\sim 10^{14}$--$10^{15}$ G) and slow
rotation ($P=2$--12 s),
and ``anti-magnetars'' \citep{gotthelf08},
which are NSs with low magnetic fields
($B\lesssim 10^{11}$ G) and moderately fast rotation, whose periods
($P\sim {\rm a\,\, few}\,\,0.1$ s) have changed
very little during their lifetime.
Also, we cannot exclude the opportunity that the CCO represents a new
type of a compact object, perhaps with an exotic composition,
 such as a strange quark star (SQS; \citealp{witten}).
 We will discuss these
options below, with account for our observational results.

\subsection{A magnetar?}
A strong support for the magnetar interpretation could be provided by
detection of pulsations with a period of a few seconds, typical for
magnetars. We found no evidence for pulsations, with an upper limit
on pulsed fraction of about 10\%--15\%, for $P>0.68$ s.
The nondetection, however, does not prove that the CCO is not a magnetar
because some of them have an even lower pulsed fraction \citep{woods}.

The shape of the CCO's
X-ray spectrum  resembles those of magnetars in the {\sl Chandra}
energy band.
For instance, the PL+BB model satisfactorily describes the spectra
of both the CCO
and magnetars. However,
according to Table 4.1 in \citet{woods},
the magnetar BB temperatures (0.4--0.7 keV)
 inferred from such fits 
are somewhat higher than the temperature of the BB component in the CCO's spectrum,
0.30--0.38 keV,
while the magnetar photon indices ($\Gamma = 2$--4) are somewhat smaller than
the photon index of the CCO's PL component, $\Gamma=
3.6$--5.2.
The fitting parameters obtained from the RCS fits are also close to
those obtained by \citet{rea} 
for some AXPs. However, the
CCO's X-ray luminosity, $L_X\sim 4\times 10^{33}$ erg s$^{-1}$,
is lower than those of magnetars, $L_X\sim 10^{34}$--$10^{35}$ erg s$^{-1}$.
It would be very interesting to compare the spectra of the CCO and magnetars
at $E>10$ keV, where
magnetars show very hard PL spectra ($\Gamma = 1$--1.5 for the total
[pulsed + nonpulsed] emission), but it 
would require hard X-ray instruments with high spatial resolution 
to separate the CCO's emission from the much brighter Cas A SNR's emission.

If the CCO were a magnetar, we would expect bursts and flares. 
However, we have not seen any significant flux variations in the
10 years of {\sl Chandra} observations. In fact, if in the last $\sim 40$
years the CCO
had produced a flare on a scale we see in SGRs,
it would 
likely has been detected by an all-sky monitor in hard X-rays
or soft $\gamma$-rays.
There was a claim by \citet{krause} 
 that the apparent motions
of Cas A filaments with tangential velocities close to the speed of light,
observed with {\sl Spitzer} at 24 $\mu$m,
could be interpreted as an infrared echo of a strong flare from the CCO
$\sim 60$ years ago,
but this interpretation has been retracted by \citet{kim} 
and 
\citet{dwek}.  

One might speculate that the CCO does possess a superstrong magnetic
field, perhaps hidden in the NS interior,
 but, being so young, it has not yet developed the properties
we see in ``mature'' magnetars. Such an interpretation is in line
with the hypothesis by \citet{bhattacharya},
 who suggest that
the superstrong magnetic field in a very young magnetar is 
originally concentrated
in the NS core, 
and it will require $\gtrsim 10^3$ yr for this field to reach the crust
and give rise to the magnetar activity. It is hardly possible to
confirm or reject this interpretation of the CCO by direct
observations (unless a burst or a flare is detected).
In the framework of this hypothesis, the X-ray emission from the
CCO should be thermal, with luminosity and temperature perhaps somewhat
higher than those of ``usual'' NSs due to additional heat that might be
released in the NS interiors by dissipation of the ``hidden'' superstrong
magnetic field, and its spectrum should be best described by an NSA model.
The observed spectrum generally does not contradict this hypothesis, 
at least if we believe that
 the NS surface temperature is nonuniform, but this would not be a unique
interpretation.
The interpretation will perhaps look more plausible if the scenario by \citet{bhattacharya}
is supported by detailed studies of established magnetars
(i.e., AXPs and SGRs).
To conclude, the magnetar interpretation of the Cas A CCO does not
seem very likely, but it cannot be ruled out, especially 
the assumption that the CCO is an ``immature'' magnetar.

\subsection{An anti-magnetar?}

In addition to the Cas A, 
CCOs in several other SNRs are known, with similar
properties \citep{pavlov2002,pavlov2004,deluca2008}.
 Those
CCOs also show thermal-like X-ray spectra, albeit with temperatures
and luminosities lower than those of the Cas A CCO, and they also do not show
pulsar or magnetar activity.
It seems natural to explore the possibility that the nature of
the Cas A CCO is the same as of
the other CCOs, just the Cas A CCO is younger.

Unlike the Cas A CCO, pulsations have been discovered in three CCOs,
with periods in the range of 0.105--0.424 s
\citep{zavlin,gotthelf05,gotthelf09}. 
Surprisingly, the upper limits on period derivatives turned out to be very 
low \citep{gotthelf07,gotthelf09,halpern},
corresponding to 
spin-down luminosities,
$\dot{E}\equiv 4\pi^2I\dot{P}P^{-3}$, lower than the X-ray luminosities
(which means that the X-ray emission is not powered by NS rotation).
To explain the fact that the corresponding lower limits on
spin-down ages are much larger than the SNR ages, $\tau_{\rm sd}\equiv P/2\dot{P}\gg
T_{\rm SNR}$,
one has to assume that the initial 
spin periods
of these pulsars were very close to their current values,
$P_0 = P (1-T_{\rm SNR}/\tau_{\rm sd})^{1/2}$.
As the upper limits on the magnetic fields\footnote{For the best-investigated
CCO, 1E\,1207.4--5209 in the PKS\,1209--51/52 SNR, the magnetic field,
$B_{\rm cycl}=6\times 10^{10}g_r^{-1}$ G
was
estimated from the absorption features \citep{sanwal} 
interpreted as
harmonics of the electron cyclotron frequency \citep{bignami}.
A very close value of the magnetic field,
$B_{\rm sd}=9.7\times 10^{10}$ G, has been recently estimated from
the period derivative determined with the aid of new {\sl XMM-Newton} 
observations 
\citep{pavlov2009b}.}, 
$B\equiv 3.2\times 10^{19}(P\dot{P})^{1/2}\lesssim\, {\rm a\,\, few}\,\times
10^{11}$ G, 
turned out to be much lower than those of magnetars (and even
of young rotation-powered pulsars),
these objects
were dubbed ``anti-magnetars'' by \citet{gotthelf08}, 
who also suggested that the Cas A CCO may be a member of this
class of NSs. 

The current data, including our new observation,
 do not contradict the assumption than the Cas A CCO 
is indeed an anti-magnetar. The nondetection of period is not 
conclusive because
the shortest period we would be able to detect is longer than the periods
of the known anti-magnetars, and the upper limit on pulsed fraction is not
low enough to exclude weak pulsation even at these long periods
(e.g., the pulsed fractions of the CCOs in the PKS\,1209--51/52 and Puppis\,A
SNRs are about 9\% and 11\%, respectively).
The Cas A CCO spectrum is well described by the same spectral models as 
those of anti-magnetars, with similar spectral parameters.
Similar to the anti-magnetars, the spectra are thermal (or at least
they contain a strong thermal component), but a single-component BB model
does not give good fits. A single-component NSA model gives acceptable
fits, but the 
radii of equivalent emitting sphere are smaller than a reasonable NS
radius. Two-component models, such as BB+BB, give very good fits, but
the NS surface layers at so high temperatures and low magnetic fields
are expected to be gaseous (plasma) atmospheres, whose spectra are 
different from BB because photons with different energies emerge
from layers with different temperatures.
Therefore, a more realistic description of the thermal emission from
anti-magnetars should be based on NSA models rather than the BB model.

As the magnetic field of an anti-magnetar is low, and the spin period is relatively
long,
accretion of SN debris soon after the 
SN explosion is possible. 
If there is even a very small fraction of hydrogen or helium
in the accreted material, 
the emission emerges from a hydrogen or helium NS atmosphere
 because of the gravitational stratification.
If the field is much lower than $10^{11}$ G, it makes
no effect on the NSA X-ray emission, so that the use of low-field NSA models
is justified. If the field is 
around $10^{11}$ G (as in 1E\,1207--5209), cyclotron absorption features are
expected in the {\sl Chandra} energy band. This would require the use of
NSA models with the cyclotron energy in the soft X-ray range (currently
unavailable in XSPEC), but the continua of such models are rather close to
the low-field NS spectra \citep{suleimanov}. 
 The relatively low fit quality and  too small size of the emitting region
in the single-component NSA fit suggests that the NS surface temperature
is not uniform. As we do not know the temperature distribution over the 
surface, we have to use the simplest approximation of two regions
with different temperatures and areas (e.g., a hot spot and a colder bulk
surface). Our fit with the NSA+NSA model has given a very good fit, 
 with
temperatures $T_{\rm eff,soft}^\infty\approx 1.6$ MK, 
$T_{\rm eff,hard}^\infty\approx
4.6$ MK, and radii $R_{\rm soft}^\infty \approx 11$ km and $R_{\rm hard}\sim
0.4$ km, for $D=3.4$ kpc.
 The bolometric luminosities of the two components are
 $L_{\rm soft}^\infty \approx 5.7\times 10^{33}$ erg s$^{-1}$ and
 $L_{\rm hard}^\infty \sim 0.4\times 10^{33}$ erg s$^{-1}$, assuming isotropic
 emission.
 Since the bolometric luminosity and the temperature of the soft component 
are consistent with the predictions of 
various NS cooling models (e.g., \citealp{tsuruta}) 
 for the Cas A CCO age,
the soft component can be interpreted as 
emission of the thermal energy stored in 
the NS interiors. However, it is not easy to
explain the origin of a small hot spot at the NS surface. 
At such low magnetic fields, the anisotropy of thermal conductivity, which
may lead to a nonuniform temperature distribution 
(e.g., the magnetic poles are hotter than the equator because
the conductivity is
higher along the magnetic field), is negligible, and, moreover, 
this effect could not produce such a small hot spot even in a
superstrong magnetic field for the dipole geometry \citep{perez}.
\citet{pavlov2000} 
 speculated that the temperature anisotropy could
be due to a nonuniform chemical composition over the NS surface (e.g., 
there might be hydrogen ``islands'', perhaps formed by accretion
onto the magnetic poles soon
after the SN explosion, on the iron surface that is colder than the
islands because the thermal conductivity of the 
degenerate NS envelope is proportional to the inverse nucleus
charge, $Z^{-1}$).
However, this interpretation looks somewhat
artificial, and it has not been supported by observational data
(e.g., heavy-element spectral lines in the CCO spectrum).
Alternatively, the hot spot(s) could be heated by ongoing 
slow accretion from a residual debris disk onto the magnetic pole(s),
but there are no observational indications, such as flux variability
or emission spectral lines, of such accretion.  
Another way to explain the origin of small hot spot(s) is to assume
the presence of a very strong toroidal magnetic field, $B\gtrsim 10^{13}$ G,
in the NS crust \citep{perez}.
 Such a field could screen the heat flux from the NS interior
toward the surface in a broad range of magnetic latitudes around the
equator, leading to a large temperature contrast between the
poles and the rest of the NS surface (see Figs.\ 8 and 9 in 
\citealp{perez}).
Although the 
toroidal field would not 
penetrate into the NS surface layers and would not affect the NS spin-down
(which is determined by a much lower poloidal [e.g., dipole] component
in this model), the hypothesis about the presence of a very strong
toroidal component in anti-magnetars may look somewhat artificial. In 
a sense, within the framework of this hypothesis, anti-magnetars
become similar to the ``immature magnetars'' in the 
scenario by \citet{bhattacharya}.

To conclude, the interpretation of the Cas A CCO as a 
young anti-magnetar
seems very plausible, but there is no an unequivocal explanation
for the nonuniformity of the surface temperature. This problem, however,
is pertinent to the established anti-magnetars, too. The anti-magnetar 
interpretation of the Cas A CCO could be confirmed by a measurement
of the CCO period with a small period derivative, which would
require dedicated observations\footnote{A 300 ks observation with the
{\sl Chandra} HRC-S detector, aimed at the search for the CCO period,
has been carried out recently (2009 March; PI D.\ Chakrabarty),
but no results have been published yet.} 
with a timing resolution 
considerably better than 0.1 s.
A firm proof of a low surface magnetic field could be provided by
 detection of two or more spectral features 
associated with cyclotron harmonics, which would require deep 
observations with an observational setup similar to that used
in our observation. To resolve the problem of nonuniform temperature
of anti-magnetars
and understand the temperature and field distributions over the NS
surface, phase-resolved spectroscopy, interpreted with the aid
of NSA models with appropriate magnetic fields, would be most helpful.

\begin{figure}
\includegraphics[scale=0.5]{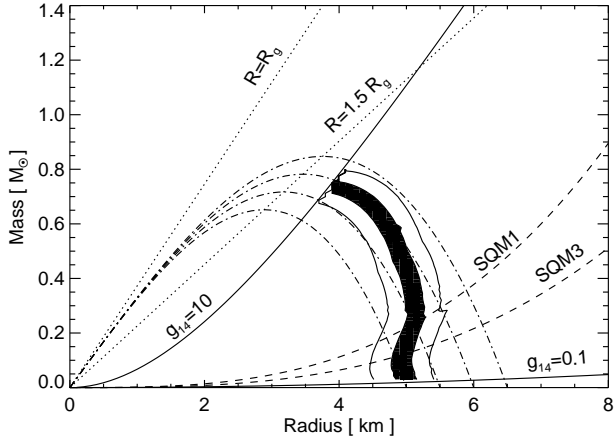}
\caption{Mass-radius region compatible with
the distance to the CCO, obtained from the {\em wabs$\times$ nsagrav} fits.
The black strip corresponds to $D=3.3$--3.5 kpc, while the white strips
leftward and rightward of the black strip correspond to $D=3.1$--3.3 and
3.5--3.7 kpc, respectively.
The dotted lines correspond to $R=R_g$ and $R=1.5 R_g$, where
$R_g$ is the gravitational (Schwarzschild) radius; the solid lines,
$g_{14}\equiv g/10^{14}\,\,{\rm cm\,s}^{-2}= 10$ and 0.1,
correspond to the maximum and minimum gravitational accelerations for which
the {\em nsagrav} models are available.
The $M(R)$ curves for two models of strange quark matter equation of state,
SQM1 and SQM3, are shown by dashed curves.
The dash-dotted curves are the loci of constant R$^{\infty}$ (note
that $R^\infty \to R$ for $M\to 0$).
}
\label{contours}
\end{figure}

\subsection{A strange quark star?}

As we have seen in \S3.2, 
 reasonably good fits to the Cas A CCO spectrum
are provided by the NSA models with a uniform effective
temperature, but these fits imply
a small emitting area, much smaller than the visible area of a NS. 
To infer the allowed range of masses and radii of the CCO,
assuming that it is covered by a low-field hydrogen atmosphere,
we have fit the observed spectrum with 
the {\em nsagrav} models on a 
grid in the $R$-$M$ plane.
Each of such fits yields the distance to the source, through the 
model normalization.
This allows one to plot the lines of constant distance
(and 
strips corresponding to ranges of distances) in the $R$-$M$ plane.
Figure 7 shows such strips around the line corresponding to $D=3.4$ kpc,
the most probable distance to Cas A \citep{reed}; 
 the black strip 
($D=3.3$--3.5 kpc) is enveloped
by the strips for $D=3.1$-- 3.3 and 3.5--3.7 kpc. We see that
the maximum allowed radius $R$ of the compact object is about 5.5 km,
while the apparent radius $R^\infty < 6.5$ km, and the object's mass
cannot exceed $0.8 M_\odot$. This $R$-$M$ domain is inconsistent with
any NS equation of state,
but it is consistent with low-mass SQS
models. The $M(R)$ curves for two equations of state of strange
quark matter, SQM1 and SQM3 \citep{lattimer}, 
shown in Figure 7, cross the $R$-$M$
domain allowed by the {\em nsagrav} fits at $R\approx 4.5$--$5.5$ km;
they correspond to the allowed mass ranges
 $M\approx 0.2$--0.3 and 0.1--0.2 $M_\odot$, respectively.
The effective temperature and bolometric luminosity,
$T_{\rm eff}^\infty \approx 2.1$ MK and $L_{\rm bol}^\infty \approx
4\times 10^{33}$ erg s$^{-1}$, do not change significantly within the
strips.

Thus, the emission from the CCO could, in principle, be interpreted as
emerging from a SQS atmosphere
which could form above
a normal-matter crust \citep{glendenning}. 
 The main problem with this interpretation is the very low star's mass.
For instance, if the SQS is formed as a result of a phase transition in a NS,
we would not expect the SQS mass to be lower than $\approx 1 M_\odot$.
One might speculate that the SQS was formed directly in the SN explosion,
but we are not aware of 
SN models that would result in such a low-mass
compact remnant.
On the other hand,  we cannot be sure that such a scenario is impossible.
Therefore, we believe that the SQS interpretation, however
exotic it looks, should not
be dismissed 
until rejected by further observations.
A possible way to check it would be very deep observations in the
near-IR (more suitable than the optical because of the
high extinction), which could detect the Rayleigh-Jeans tail of thermal
emission from the bulk of the NS surface. If the small region associated
with the CCO X-ray emission is, in fact, just a heated area on the
NS surface, while the rest of the surface is substantially colder,
the near-IR observation could detect this colder emission and prove
that its size is much larger than that seen in X-rays.
The current deepest near-IR limit, inferred from
{\sl HST} NICMOS observations, is $H>24.6$ \citep{fesen2006},
   but 
the emission from the NS surface is expected to be even fainter, 
i.e., deeper observations are required to detect it.
The SQS interpretation would also be 
called into question if future X-ray
observations show this object to be an anti-magnetar. (Although
anti-magnetars might, in principle, be SQSs, the more conventional
NS interpretaion is preferable
until proven otherwise, according to the Occam's razor.)

\subsection{Summary}
To summarize, we believe that the anti-magnetar interpretation of
the Cas A CCO,
similar to the three other CCOs proven to be anti-magnetars, 
is currently the most plausible. 
In this interpretation,
the CCO's emission
with the bolometric luminosity 
$L_{\rm bol}^\infty \sim 6\times 10^{33}$ erg s$^{-1}$ 
emerges from a NS atmosphere with
 nonuniform effective temperature,  relatively low magnetic field, $B\lesssim
10^{11}$ G, and  very slow spin-down.
The origin of
the temperature nonuniformity is not clear, as well as for
the other anti-magnetars;
 it might be caused by a much stronger toroidal magnetic field in the
NS crust. The most direct way to check the anti-magnetar
interpretation is to
measure the CCO's period and period derivative and/or detect
spectral features that could be interpreted as harmonics of the
electron cyclotron frequency.

There is no observational evidence
of the Cas A CCO being a magnetar similar to the
currently known SGRs and AXPs. We, however, 
cannot rule out the possibility that it is an immature magnetar,
whose ultrastrong magnetic field is hidden in the NS interiors
and does not make a strong effect on the observable properties.
If this is the case, the Cas A CCO may eventually turn into an 
``ordinary'' magnetar, but we consider this hypothesis rather
speculative.

Finally, we cannot firmly reject the possibility that the Cas A
CCO is a SQS, with a radius of $\approx 5$ km and a mass $\lesssim
0.8 M_\odot$, covered by a normal-matter crust and an atmosphere
with $kT_{\rm eff}^\infty \approx 0.18$ keV and $L_{\rm bol}^\infty
\approx 4\times 10^{33}$ er s$^{-1}$. It remains unclear
how the SQS with so low mass could be formed. Future observations will
help check this interpretation.

\acknowledgments
We thank Slava Zavlin for calculating {\em nsagrav} models for a broader
parameter domain, Dany Page for providing the NS and SQS
 $M(R)$ relations for various
equations of state, Leisa Townsley for the discussions of the CTI effects
in ACIS, Nanda Rea for the discussion of the RCS spectral model,
Sandro Mereghetti for the discussion of the {\sl XMM-Newton} observation
of Cas A,
and Oleg Kargaltsev, Zdenka Misanovic and Bill Joye for their useful advice
on the data analysis.
Support for this work was provided by the National Aeronautics and Space Administration through {\sl Chandra} Award Number GO6-7055X issued by the {\sl Chandra} X-ray Observatory Center, which is operated by the Smithsonian Astrophysical Observatory for and on behalf of the National Aeronautics Space Administration under contract NAS8-03060. The work was also partially supported by NASA grant
NNX09AC84G.

\input{table}

\end{document}

%% file: table.tex

\begin{deluxetable}{lcccccccc}
\tablewidth{0pt}
\tablecaption{Fitting parameters for different spectral models}
\tablehead{
\colhead{Model} & \colhead{$N_{\rm H,22}$\tablenotemark{a}} & \colhead{$\Gamma$\tablenotemark{b}} &
\colhead{$kT^\infty$\tablenotemark{c}} & \colhead{${\mathcal N}_{-2}$\tablenotemark{d}} & \colhead{$R^{\infty}/D$\tablenotemark{e}} &
\colhead{$F_{-13}$\tablenotemark{f}} & \colhead{$F_{-12}^{\rm un}$\tablenotemark{g}} & \colhead{$\chi_\nu^2$ (dof)} }
\startdata 
{\em wabs}$\times$PL & $2.81^{+0.13}_{-0.13}$ & $5.23^{+0.17}_{-0.16}$ & ... & $1.38^{+0.29}_{-0.23}$ & ... & $6.68_{-0.17}^{+0.17}$ & 35.58$_{-6.85}^{+8.99}$& 1.21(125) \\

{\em wabs}$\times$BB & $1.32^{+0.08}_{-0.07}$ & ... & $402^{+11}_{-11}$ & ... & $0.91^{+0.09}_{-0.07}$/3.4 & $6.53^{+0.17}_{-0.17}$  & 1.84$_{-0.12}^{+0.14}$ & 1.54 (125) \\

{\em wabs}$\times${\em nsa}\tablenotemark{h} & $1.57^{+0.08}_{-0.10}$ & ... & $185^{+9}_{-8}$ & ... & 13.06/$8.5^{+1.6}_{-1.0}$ & $6.62^{+0.17}_{-0.17}$ & 2.47$_{-0.20}^{+0.24}$ & 1.24 (125) \\

{\em wabs}$\times${\em nsagrav}\tablenotemark{i} & $1.57_{-0.08}^{+0.09}$ &...& $182_{-8}^{+8}$ &...& $5.49_{-0.64}^{+0.74}$/3.4 & $6.61_{-0.14}^{+0.15}$&2.81$_{-0.32}^{+0.23}$&1.25 (125) \\

{\em vphabs}$\times${\em nsa}\tablenotemark{j} & $1.46^{+0.10}_{-0.08}$ & ... & $190^{+4}_{-13}$ &...& 13.06/$8.5^{+1.1}_{-1.4}$ & $6.62^{+0.17}_{-0.17}$ &2.82$_{-0.33}^{+0.26}$ & 1.19/(124) \\

{\em wabs}$\times$RCS\tablenotemark{k} & 1.64$^{+0.07}_{-0.06}$ & ...  & 283$^{+25}_{-180}$& ... & ... & 6.69$_{-0.17}^{+0.18}$ &2.80$_{-0.24}^{+3.64}$& 1.11(124) \\ 

{\em wabs}$\times$(PL+BB)& $2.27_{-0.49}^{+0.36}$ & $4.62_{-1.07}^{+0.56}$ & ...               & $0.44^{+0.55}_{-0.36}$ &  .... & $6.70_{-0.17}^{+0.17}$ &9.96$_{-6.14}^{+8.34}$& 1.12 (123)\\
                      ...& ...                    &  ...                    & $332_{-30}^{+52}$ &...                  & $1.19^{+0.46}_{-0.57}/3.4$ &...& 1.37$_{-0.25}^{+0.31}$&...\\

{\em wabs}$\times$(BB+BB)&1.71$_{-0.17}^{+0.21}$&...&305$_{-38}^{+35}$&...&1.92$^{+0.97}_{-0.51}$/3.4&6.68$_{-0.18}^{+0.17}$&2.45$_{-0.37}^{+0.49}$&1.11(123)\\
...&...&...&614$_{-90}^{+170}$&...&0.19$_{-0.11}^{+0.15}$/3.4&...&0.45$_{-0.12}^{+0.18}$&...\\

{\em wabs}$\times$({\em nsa}+{\em nsa})&1.81$_{-0.80}^{+0.81}$&...&141$_{-35}^{+27}$&...&13.06/4.2$_{-2.4}^{+1.7}$&6.68$_{-0.17}^{+0.17}$&3.24$_{-0.53}^{+0.69}$&1.11(123) \\
...&&...&394$^{+92}_{-86}$&...&13.06/119$_{-85}^{+180}$&...&0.28$_{-0.12}^{+0.23}$&... \\


\enddata
\tablecomments{The fits are for the source spectrum extracted from the 
$1.476''$ radius aperture, background taken from the 
$2.26''$--$3.94''$ annulus. The errors of fitting parameters are given at the 90\% confidence level for one interesting parameter. }
\tablenotetext{a}{Hydrogen column density in units of $10^{22}$ cm$^{-2}$.}
\tablenotetext{b}{Photon index for PL spectra.}
\tablenotetext{c}{Effective temperature as seen by a distant observer, in units of eV. }
\tablenotetext{d}{PL normalization = photon spectral flux at 1 keV, in units of $10^{-2}$ photons cm$^{-2}$ s$^{-1}$ keV$^{-1}$.}
\tablenotetext{e}{Radius (as seen by a distant observer) in units of km, and distance in kpc. In the {\em nsa} model, the NS radius and mass are fixed at $R=10$ km
and $M=1.4 M_\odot$ ($R^\infty = 13.06$ km), while the distance is the fitting parameter. }
\tablenotetext{f}{Observed (absorbed) energy flux in the 0.6--6 keV band, in units of $10^{-13}$ ergs cm$^{-2}$ s$^{-1}$, calculated with the {\em cflux} model in XSPEC.}
\tablenotetext{g}{Unabsorbed energy flux in the 0.6--6 keV band, in units of $10^{-12}$ ergs cm$^{-2}$ s$^{-1}$, calculated with the {\em cflux} model in XSPEC. The corresponding luminosity
is $L_X = 1.38\times 10^{33} F_{-12}^{\rm un}$ erg s$^{-1}$, for $D=3.4$ kpc. }
\tablenotetext{h}{NSA model for fixed $M=1.4 M_\odot$ and $R=10$ km.}
\tablenotetext{i}{The fit is for fixed $M=0.25 M_\odot$. The distance (normalization) was fixed, while the radius $R$ is a fitting parameter ($R=5.08^{+0.68}_{-0.59}$ km).}
\tablenotetext{j}{The result is for variable Si abundance; its fitted value is $3.2^{+1.3}_{-1.3}$.}
\tablenotetext{k}{Fit with the resonance Compton scattering model, for fixed $\tau_{\rm res}=1.0$. In addition to $N_{\rm H}$ and $kT$, the fitting parameters are 
$\beta_T=0.33^{+0.02}_{-0.09}$ 
and {\em norm} = 
$0.7_{-0.5}^{+100}\times 10^{-4}$.} 
\label{table:spectra}
\end{deluxetable}